 \documentclass[final,5p,times,twocolumn,authoryear]{elsarticle}

\usepackage[utf8]{inputenc}
\usepackage{lipsum}
\usepackage{xcolor}

\usepackage{pdflscape}
\usepackage{array}
\usepackage{arydshln}

 \usepackage{cite} 
 \usepackage{epstopdf} \usepackage{amsmath,amssymb,nicefrac} \usepackage{psfrag} 
 \usepackage{subfigure}
 \usepackage{url} 
 \usepackage{multirow}
 \usepackage{rotating} 
 \usepackage{pifont} 
 \usepackage{boxedminipage} 

 \usepackage{bbm} 
 \usepackage{float} 
 \usepackage{lscape} 
 \usepackage{amsfonts} 
 \usepackage{helvet} 
 \usepackage{amsxtra} 
 \usepackage{multirow} 
 \usepackage{titletoc} 
 \usepackage{algpseudocode} 
 \usepackage{lastpage} 
 \usepackage{verbatim} 
 \usepackage{graphicx}
\usepackage[toc,page]{appendix}

\usepackage{cite} 
\usepackage{hyperref}
\usepackage{tensor}
\hypersetup{
	colorlinks=true,       
	linkcolor=red,        
	citecolor=blue,        
	filecolor=magenta,     
	urlcolor=blue
}

\journal{New Astronomy}

\begin{document}

\begin{frontmatter}


 \tnotetext[a]{Correspondence:}
\author{Shad Ali\corref{cor1}\fnref{add1}}
 \ead{shad.ali88@yahoo.com}
  \author{Tong Liu\corref{cor1}\fnref{add1}}
   \ead{tongliu@xmu.edu.cn}

\title{The CR Volume for Black Holes and the Corresponding Entropy Variation: A Review}


\affiliation[add1]{organization={Department of Astronomy},
            addressline={Xiamen University}, 
            city={Xiamen},
            postcode={361005}, 
            state={Fujian},
            country={China}}

\begin{abstract}
    This paper reviews the work done on black hole interior volume, entropy, and evaporation. An insight into the basics for understanding the interior volume is presented. A general analogy to investigate the interior volume of a black hole, the associated quantum mode's entropy, and the evolution relation between the interior and exterior entropy is explained. Using this analogy, we predicted the future of information stored in a BH, its radiation, and evaporation. The results are noted in tables (\ref{tab:1}) and (\ref{tab:2}). To apply this analogy in BH space-time, we investigated the interior volume, entropy, and evaluation relation for different types of BHs. Finally, we also investigated the nature of BH radiation and the probability of particle emission during the evaporation process.
\end{abstract}




\end{frontmatter}




\section{Terminology}\label{terminology}
\subsection{Black hole interiors} \label{BH-int}
A Black Hole (BH) is a region of space-time with intense gravity such that nothing could escape from its surroundings \citep{Wald:1984rga, Weber, Carroll:2004st, Stephani:2004ud, Misner:1973prb}. According to Hawking's uniqueness theorem \citep{Hawking:1971vc}, all BHs belong to the Kerr Newman BH family that is characterized by its mass $m$, charge $q$, and angular momentum $J$ \citep{Mazur:2000pn, Saida:2011wj, Hawking:1976de}. Its horizon behaves like a trapped region where both the future-directed and null geodesics are orthogonal to it but converging due to strong gravity i.e., the outgoing light is also dragged inward. The horizon is a cut-up region between its interior and the surrounding space. The study of the BH interior has challenging issues due to the interchanging space-time coordinates across the horizon \citep{Hawking:1994ss}.  The cosmic Censorship hypothesis work of Roger Penrose has provided insight into BH interiors \citep{Penrose:1999vj}. An important point of this work is the inner horizon of a BH (charged, rotating, or both) should be unstable. Recently a deep concern has been made regarding these hypotheses in using mathematical relativity, semi-classical gravity, and numerical studies.  
\citep{Penrose:1971uk, Simpson:1973ua}.

\subsection{Hyper-surface}\label{hyper-surface}
Generally, a hyper-surface is the generalization of an ordinary $ 2-$dimensional surface embedded in $3-$dimensional space to an $(n-1)-$dimensional surface embedded in an $n-$dimensional space \citep{Christodoulou:2014yia, Christodoulou:2016tuua, Hsu:2007dr}. As the investigation of interior volume in flat space-time is not the same as in curved space-time, thus it is important to understand the meaning of hyper-surface in both flat and curved space-time \citep{Misner:1973prb}.

\subsubsection{In flat space-time}\label{flat-spce-time}
Mathematically, the interior volume $V$ bounded by a sphere $S$ of radius $R$ immersed in flat Minkowski space-time is $\frac{4}{3}\pi$ times its cubic radius \citep{Dolan:2013ft, Wang:2021llu, Johnson:2019wcq}. At the same time, its surface area is $\pi$ of times its squared radius. It is a $3-$dimensional space-like spherically symmetric hyper-surface $(say \sum)$ bounded by that sphere $S$ having volume $V$ \citep{Estabrook:1973ue, Cordero-Carrion:2001jpf}. It is the maximal volume determined by the largest $\sum$ in the interior of sphere $S$. So, to define the interior volume bounded by a sphere $S$ in Minkowski space-time, one should choose the largest $\sum$ in the interior $S$. This  $\sum$ must satisfy two primary conditions:

    \begin{itemize}
        \item the simultaneity condition, and 
        \item It must be the largest space-like spherically symmetric hyper-surface bounded by sphere $S$.
    \end{itemize}
    
\begin{figure}[!ht]
	\centering 
	\includegraphics[width=0.3\textwidth, angle=0]{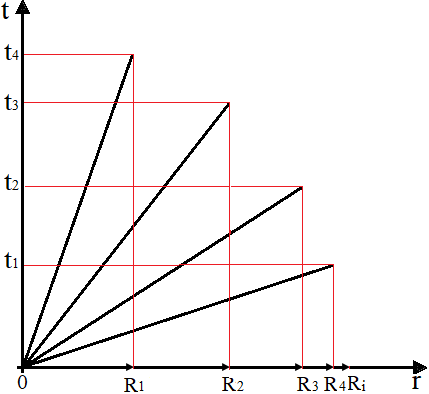}	
	\caption{Choosing the maximal hyper-surface in flat Minkowiski space-time with $t=0$ results in the maximal interior volume of the sphere $S$. This figure shows that any contribution to the time-like direction leads to a reduction of interior volume.} 
	\label{image-1}%
\end{figure}

Both conditions could be equivalently treated in Minkowski's space-time \citep{dInverno:1992gxs, Weber, Carroll:2004st}. Consider a sphere $S$ with coordinates $R^2=x^2+y^2+z^2$ at $t=0$. A spherically symmetric hyper-surface bounded by a sphere $S$ can be defined as $t=t(r), ~ r \in[0, R]$, and $t(R)=0$ and the interior volume bounded by this hyper-surface is  
 \begin{equation}\label{V1}
 V=4\pi\int_0^R [r^4(1-[\partial_rt(R)]^2)]^\frac{1}{2} dr,
 \end{equation}
This equation shows the maximum volume bounded by the largest space-like hyper-surface at $t(R)=0$ is $V=\frac{4}{3}\pi R^3$ \citep{Padmanabhan:2002sha, Parikh:2005qs}. The same case can be understood by analyzing Fig. (\ref{image-1}), where for an inertial observer, the simultaneity surfaces are straight lines in a $t-r$ plane. 

From the above two conditions, let us consider $R$ is the radius of the largest hyper-surface at time $t(R)=0$, then we can write the metric component as $g_{\mu v}=(0,~R)$, and the metric is $ds^2=dR^2$ so, its maximal volume is the same as $V=\frac{4}{3}\pi R^3$. Now, let us consider $R _i$ as another hyper-surface at $t(R)=t_i$ with $i=1,2,3,4$, then the metric component $g_{\mu v}=(-t(R_i),~ R_i)$, the metric is $ds'^2=dR_i^2-dt(R_i)^2 $ that gives the volume $ V_i$. On comparison, we will get the volume as $V> V_i$, and hence $ds^2 > ds'^2$. Note that any spherically symmetric sphere can reside only one largest $\sum$ inside it.  The inertial frame will be defined for the $\sum$ bounding of the largest volume (also called the proper volume). Subsequently from Eq. (\ref{V1}), we can say that any contribution to the time-like direction can reduce interior volume.

\subsubsection{In curved space-time}\label{curved-space-time}
\begin{figure}[!ht]
	\centering 
	\includegraphics[width=0.45\textwidth, angle=0]{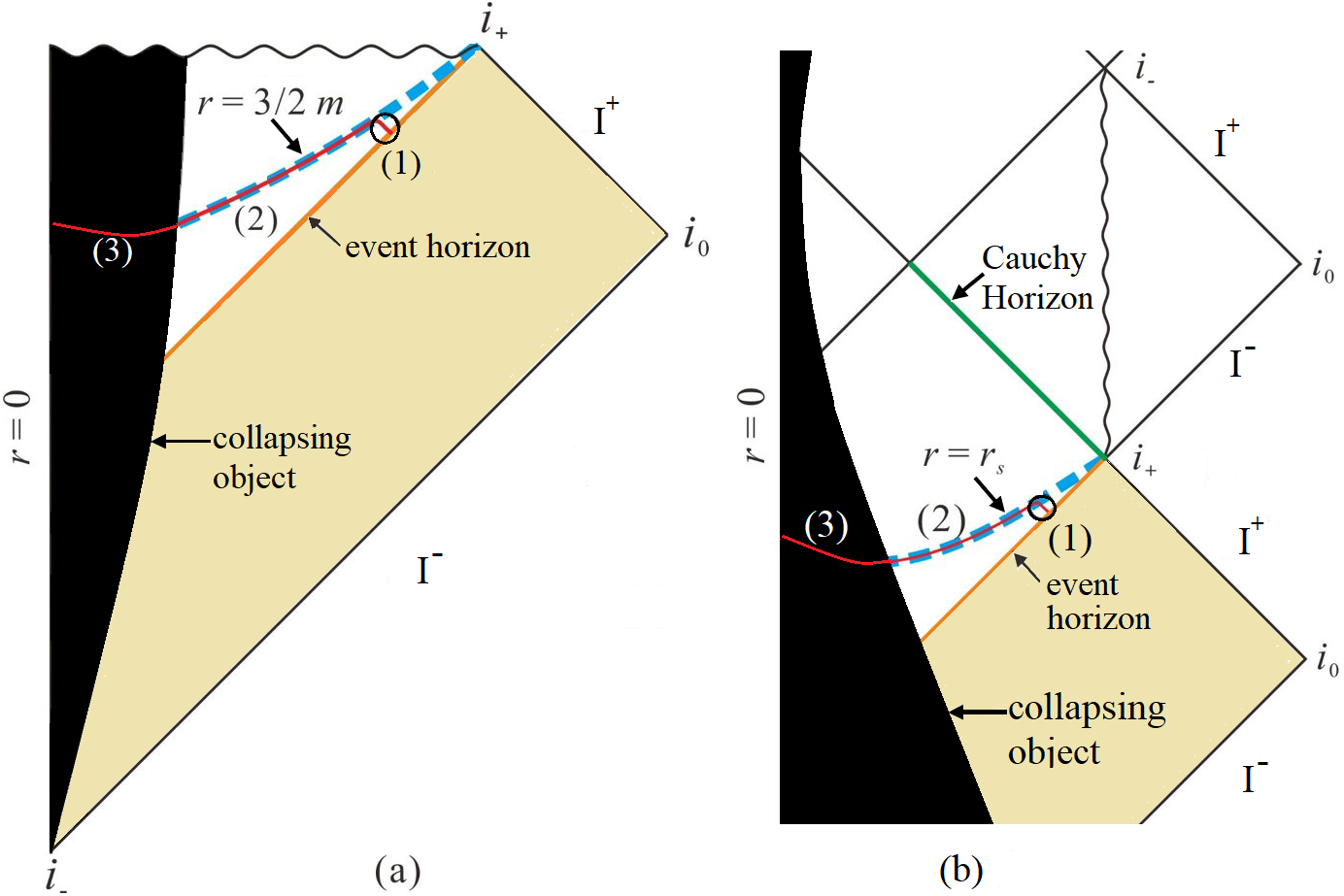}	
	\caption{(a) The largest hyper-surface (red line) drawn from the horizon to the center of a collapsed object. Part $(1)$ is the null part, part $(2)$ is the stretched segment between Part $(1)$ and the boundary of the collapsed surface, and part $(3)$ is the collapsed part with a finite volume. (b) A Penrose diagram of Kerr's geometry showing the largest hyper-surface (red line) and its segments labeled by $(1), ~(2), ~\rm and ~(3)$.} 
	\label{image-2}%
\end{figure}

In curved space-time, the topic of BH interior volume is not simple as that in Minkowiski space-time \citep{Carroll:2004st, Mann:2015luq, Parker:2009uva}. It demands different techniques to be adopted for choosing the largest space-like hyper-surface.  Firstly, Christodoulou and C. Rovelli (CR) \citep{Christodoulou:2014yia, Christodoulou:2016tuua} investigated the maximal interior volume of BH after defining the largest hyper-surface in the interior of BH. According to their work, the largest hyper-surface bounded by an $n-$sphere can be explained by a Penrose diagram as shown in Fig. (\ref{image-2}). In this figure,  a space-like curve is drawn (from the horizon to the center of the collapsed object) and is divided into three parts labeled as $(1)$, $(2)$, and $(3)$. Section $(1)$ connects the hyper-surface to the horizon and is a null part of this hyper-surface. The long stretch part of the hyper-surface (i.e., section $(2)$) has a nearly constant radius (say $r_v$). According to CR work, the main contribution to the interior volume of a BH comes from section $(2)$, and section $(3)$ is part of the hyper-surface that connects the long stretched part to the center of the collapsing object i.e., $(~ r = 0)$. Let the volume bounded by each part be $V_1,~ V_2$, and $V_3$. By this terminology, $V_1=0$, and $V_3$ is the volume of the collapsed part. Since the space-time inside a collapsing object acquires a time-like killing vector field that is why it has a finite volume. Hence, $V_3=constant$ (will give a finite contribution to the interior volume of the BH). Thus, the total contribution of these two parts at the longest Eddington time $(v)$ is finite and can be ignored. This means that the total contribution to the BH's interior volume comes from $V_2$. The total volume bounded by the largest hyper-surface is $V$ which increases linearly with $v$ due to $V_2$. So, we can take part $(2)$ as the largest space-like hyper-surface with $r=r_v$. Hence, the interior volume bounded by the BH at $r=r_v$ will be its maximum interior volume $V$.

\subsection{Advanced time or Eddington time (\textit{v}) and Space-time metric}\label{adv-time}
The metric on an arbitrary hyper-surface is generally written as \citep{Chandrasekhar:1985kt, Hawking:1973uf, Wald:1984rga, Weber, Carroll:2004st, Stephani:2004ud, Misner:1973prb} 
\begin{equation}
    ds^2=g_{\mu v}dx^{\mu}dx^{v},
\end{equation}
here $g_{\mu v}$ is the space-time metric. Due to its quadratic character, it is always easy to diagonalize by coordinated transformation. For example, the Schwarzschild metric case 

$$ds^2=-g_{00}dT^2+h_{ij}dx^idx^j$$
\begin{equation}
=-\left(1-\frac{r_s}{r}\right)dt^2+\left(1-\frac{r_s}{r}\right)^{-1}dr^2+r^2\left(d\theta^2+sin^2\theta d\phi^2\right),
\end{equation}

there is an artificial curvature singularity at $r=r_s$ due to the bad selection of space$-$time coordinates. To avoid this singularity one needs to choose the Eddington-Finkelstein coordinates where the time coordinate (Eddington time) is replaced as
\begin{equation}\label{Edd}
v:=t+r^*=t+ \int{\frac{dr}{f(r)}}=t+r+r_s ln(r-r_s),
\end{equation}
\begin{equation*}
    f(r)= 1-\frac{2M}{r},
\end{equation*}
and the new Schawarzschild metric becomes
\begin{equation}\label{EDmet+T}
{ds}^2=-dT^2+\left(-f(r) \dot{v }^2+2 \dot{v } \dot{r}\right)d \lambda ^2+r^2d\Omega^2,
\end{equation}
here 
$$g_{TT}=-1,\quad g_{\lambda \lambda }=\left(-f(r)\dot{v }^2 +2 \dot{v } \dot{r}\right), \quad g_{\theta \theta }=r^2 $$
$$g_{\phi \phi }=  r^2 \sin ^2 \theta,$$

It is a time-dependent equation so, the interior of the BH is not static, i.e. for any constant radial component, the hyper-surface is variable at the given time $T$. Moreover, as these investigations are made for $v>>M$, and $r=\frac{3}{2}M$ near the largest hyper-surface as $T$ increases, the proper time between two neighboring hyper-surfaces must tend to be zero and there will be no evolution for the BH. This means that the statistical quantities will not be affected by the time-dependent character of the metric. In our investigations, $T$ is approximately constant. That is why the space-like hyper-surface leads to the CR volume. Next, we follow the statistical way to find the entropy of the quantum in the scalar field in Schwarzschild's BH. 
\begin{equation} \label{collapsedmass}
ds^2=(-f(r) \dot{v}^2+2\dot{v}\dot{r})d\lambda^2+r^2 (d\theta^2+sin^2\theta d\phi^2),
\end{equation}
These coordinates are acceptable for investigations in the interior of a BH. In the case of rotating BHs \citep{Wald:1984rga, Weber, Carroll:2004st, Wiltshire:2009zza, Narlikar:1986kr}, the Eddington form can be written as 
\begin{equation}\label{Kerrmet}
\begin{aligned}
ds^2=-\frac{\Delta -a^2   \sin ^2\theta}{\rho ^2}dv^2+\rho ^2 d\theta ^2+2 dv dr \\ +\frac{A^2  \sin ^2\theta}{\rho ^2}d\phi^2-2 a  \sin ^2 \theta dr d\phi -\frac{4 a   M r^2 sin\theta^2}{\rho ^2}dv d\phi,
\end{aligned}
\end{equation}
where $\Delta$, $\rho$, $A$, and $J$ have conventional meanings as in literature. Similarly, the metric in lower space-time dimensions (like BTZ BHs) is \citep{Banados:1992wn, Carlip:1995qv, Emparan:2020znc, Gukov:2003na}

\begin{equation}\label{E-Fmetric}
ds^2=-f(r) dt^2+\frac{dr^2}{f(r)}+r^2 (N^\phi (r) dt+d\phi)^2,
\end{equation}
here, $f(r)$ is the lapse function, and $N^\phi$ is the shift function defined as 
\begin{equation}
N^\phi (r)= -\frac{J}{2r^2},  \left (\left | J \right |\leq ml  \right ),
\end{equation} 
The Eddington form of this Eq. (\ref{E-Fmetric}) can be written as 
\begin{equation}
    ds^2=(-f(r) \dot{v}^2+2\dot{v}\dot{r})d\lambda^2+r^2 (N^\phi (r) dt+d\phi)^2,
\end{equation}
here, $l$ is the AdS radius of BH (related to cosmological constant $\Lambda=-\frac{1}{l^2}$), $\phi$ is the period ranging from $0$ to $2\pi$, $J$ is the angular momentum related to angular velocity $\Omega (r)$ and $m$ is the AdS mass. For $J=0$ (static BH), we get 
\begin{equation}\label{metric}
ds^2=-f(r) dt^2+\frac{dr^2}{f(r)}+4r^2 d\phi^2=(-f(r)\dot{v}^2+2\dot{v} \dot{r}) d\lambda^2+4r^2 d\phi ^2,
\end{equation}
Some other geometries of space-times metrics in the Eddington coordinate can also be found in the literature e.g. \citep{Ong:2015tua, Ali:2019icq, Ali:2018sqk}. Note that the actual meaning of Eddington's time remains the same as the definition above (\ref{Edd}). 

\subsection{Black hole interior volume}\label{int-vol}

The main purpose of the BH study is to understand the nature of BH which is up to date a mystery in one way or another. Many attempts are made to probe its full structure and properties. Among these, the interior volume is a factor probed by many authors e.g., \citep{Grumiller:2005zk, DiNunno:2009cuq, Ballik:2010rx, Cvetic:2010jb, Gibbons:2012ac, Ballik:2013uia, Finch:2012vli, Iliesiu:2021ari, Chew:2020twk, Davidson:2010xe}. Such an attempt was made in 2015 by M. Christodoulou and C. Rovelli (CR) \citep{Christodoulou:2014yia} to solve the problem of BH interior volume by choosing the largest space-like hyper-surface in the interior of a spherically symmetric BH.  They considered the BH formed under the collapsed process (\ref{collapsedmass}) so, by using Eddington Finkelstein coordinates they defined the interior volume of BH.

Using CR's notion, this work was extended to the rotating BH in Ref. \citep{Bengtsson:2015zda}, to charged rotating BH \citep{Wang:2019ear, Haldar:2023pcv, Biro:2019rms}, RN BH case is discussed in \citep{Han:2018jnf, Wang:2018txl, Haldar:2019buj}, quasi-static spherically symmetric charged BH \citep{Jiang:2020rxx}, the BTZ BH in Refs. \citep{Zhang:2019pzd, Ali:2020qkb}, its Lagrangian formalism are demonstrated in \citep{Maurya:2022vjd}, divergent volume \citep{Zhang:2020gbv}, non-commutative BH \citep{Zhang:2016sjy} and many others.

\subsection{Black hole exterior and interior Entropy}\label{int-ext-entropy}

In thermodynamics, the term entropy refers to the microscopic and macroscopic connections of a system e.g., the entropy of a gas is the microscopic heat transfer to the available number of micro-states for the gas molecules \citep{Jacobson:2005kr, Frolov:2018awz, Bekenstein:1973ur, Bekenstein:1972tm, Wald:1999vt, Wald:2002mon, Srednicki:1993im, Frampton:2008mw, Egan:2009yy, Bhaumik:2016sav}. Similarly, the BH entropy is a test for unifying gravitational and quantum mechanical theories \citep{tHooft:1984kcu}. According to Bekenstein and Hawking \citep{Bekenstein:1973ur, Hawking:1975vcx}, the BH is a thermodynamical object having entropy which is irreversibly related to its surface area. This entropy is called Horizon entropy or Hawking entropy denoted by $S_{BH}$   \citep{Bekenstein:1972tm, Bekenstein:1973ur, Almheiri:2020cfm, Strominger:1996sh}. Later, the four laws of BH thermodynamics are proposed in Ref. \citep{Bardeen:1973gs}. These investigations started a new dialogue of BH thermodynamics as a wide research area in BH's Physics. Several techniques in connection with the thermodynamics of BH were presented to explain its structure and properties. Due to the mysterious nature of  BH, the thermodynamics claim Hawking raised the questions of radiation and information loss which is termed an information paradox. It is suggested, that the entropy of the thermodynamics system is related to the information contained in the BH system \citep{Marolf:2017jkr}. So, by finding the correct entropy one can solve the problem of the information loss paradox. In search of these facts, Baocheng Zhang followed CR's investigations and found the entropy for the quantum modes of the scalar field. Both the interior volume and associated entropy are proportional to Eddington time \citep{Zhang:2015gda}. It means that the entropy variation in the interior of a BH may affect the statistical quantities. This is the main point to consider for understanding the BH interior. This point has been studied by several authors and found that the BH entropy variation can lead to understanding the interior of BHs with the evaporation and the information loss paradox \citep{Hawking:1974rv, Hawking:1996ny}.

\section{Introduction\label{introduction}}

A BH has a horizon making it a precise object compared to those we can see in our surroundings. The existence of the BH horizon is the main reason for the interchange of coordinates across the BH horizon \citep{Penrose:1969pc, Rindler:1956yx, Ng:1993jb} and creates the problem of the information paradox \citep{Hawking:1976ra}. After introducing the BH thermodynamics and its horizon the main question was "\textit{What is the size of a BH? How is the interior of BH and what is the future of an object entering BH's boundary?}" Amongst many attempts for these questions, Parikh discussed the interior volume of BH using stationary space-time \citep{Parikh:2005qs}. His proposed volume was independent of  time coordination. It means that his proposed volume for the interior of BH was constant in time. As for the information exchange from BH to its surroundings, the Bekenstein and Hawking entropy needs the BH's internal volume to be time-dependent (i.e., variable). The main point for understanding the BH interior is a space-time with a finite horizon area and infinite volume. When the radius goes to infinity the horizon must be constant. If this case is possible, one could construct the largest hyper-surface that bounds the maximal interior volume inside the horizon. In Ref \citep{Grumiller:2005zk}, the BH volume was found proportional to BH's area. DiNunno and Matzner \citep{DiNunno:2009cuq} suggested that for the volume inside a BH, one needs to define particular three-space coordinates to evaluate the interior volume \citep{Caticha:2005qd}. These three-space coordinates may be explicitly time-dependent but have a limit of the integral to compute this volume. A similar $\rm3d$ space coordinates can be constructed about a rotating BH (that must be axisymmetric rather than spherically symmetric). Any definition of time in a rotating BH space-time leads to the possibility of evaluating the volume bounded by the horizon. Ref. \citep{Cvetic:2010jb}, treating the cosmological constant (Pressure) as a dynamical variable in the first law of BH thermodynamics, and  the thermodynamics volume was calculated for static multi-charge solutions in four, five, and seven-dimensional gauged supergravities \citep{Ashtekar:1984zz}; rotating Kerr-AdS BHs in arbitrary dimensions \citep{Gibbons:2004uw}; and certain charged rotating BHs in four and five-dimensional gauged supergravities \citep{Myers:1986un} etc. For non-rotating BHs \citep{Tangherlini:1963bw},  the thermodynamical volume was claimed to be an integral of the scalar potential over the interior volume of BH. In contrast, for rotating BHs, the thermodynamics volume and the geometric volume differ from each other by a shift related to the angular momenta of the BH. Similarly, several other investigations are made to evaluate the interior volume of BH in such a way that it could satisfy the interior characteristics of a BH  \citep{Gibbons:2012ac, Ballik:2013uia, Kawai:2015uya}. 

Generally, the volume can be defined as 
\begin{equation}
V=\int_{S^n}{\sqrt{g}dr d\theta d\phi},
\end{equation}
where $g$ is the determinant of metric and $S^n$ represents an $n-$sphere. A $4-$dimensional flat BH can be regarded as an $n=2$ sphere. By determining $g$, Parikh \citep{Parikh:2005qs} found a constant volume of BH. Similarly \citep{DiNunno:2009cuq} found the volume of a stationary BH for $t=0$ as the null volume at the horizon. Due to the existence of BH's entropy and temperature \citep{Hawking:1975vcx, Bekenstein:1973ur, Bardeen:1973gs}, these investigations of BH interior volume using old techniques led the researcher to an unsatisfactory result about the BH's constant volume. In Ref. \citep{Christodoulou:2014yia, Christodoulou:2016tuua}, Christodoulou and Rovelli  considered Schawarzschild BH formed under a collapsed process with the largest hyper-surface bounding the maximal interior volume and found that the interior volume increases linearly with advanced time $(v)$. 

From their numerical analysis, the main contribution to the interior volume of BH was from the central part of the space-like hyper-surface at the radius $r=r_v$ as shown in Fig.(\ref{image-2}). They  showed that the interior volume of a space-like spherically symmetric hyper-surface in the interior of the BH is given by

\begin{equation}
V=\int ^v \int_{S^2} Max\left[-r^4 f(r)\right]^{\frac{1}{2}}d\theta d\phi dv,
\end{equation}

the factor $Max\left[-r^4 f(r)\right]^{\frac{1}{2}}$ is maximized for some value of $r=r_v$ to get
\begin{equation}
    V_{\sum}=4\pi \sqrt{-r_v^4 f(r_v)}v,
\end{equation}
here calculating $r_v$, one can easily get the interior volume of BH as a function of Eddington time $(v)$. Later, these investigations of Christodoulou and Rovelli are extended by many authors by using different BHs space-times with surprising results as noted in the table (\ref{tab:1}) below. In reference \citep{Yang:2018arj} the rate of mass change is taken into account, and it is found that volume increases toward evaporation. Whereas, in high-dimension cases, the increase in the rate of mass change decreases with the number of dimensions.
The proportionality relation between BH interior volume with $v$ distinguishes these investigations from those of earlier ones. This property could also be witnessed for change of statistical quantities of quantum fields in the BH's interior and is expected to solve the problem of the BH information paradox as claimed by Parikh \citep{Parikh:2005qs}.  As the BH information is associated with its entropy, Baocheng Zhang first discussed the issue of the BH information paradox by using the quantum mode entropy of Schwarzschild's BH \citep{Zhang:2015gda}. 

Baocheng Zhang proposed the largest hyper-surface in the interior of BH and calculated the total number of quantum states $g(E)$ contained in the massless scalar field by using the Klein-Gordon equation \citep{Zhang:2015gda, Cordero-Carrion:2001jpf}

\begin{landscape}
\begin{table}[!th]
\caption{Results from Maximal hyper-surface and its bounded interior volume for different BHs.}
\label{tab:1}
\begin{tabular}{||c|p{3.5cm}|p{6cm}|p{3cm}|c|| }
\hline
 \hline
  S. No. &Black hole  & \qquad $r_v$ &The numerical values of $r_{v}$& $V_{\sum}$ \\ 
 \hline
 \hline
  1. & Schawarzschild \citep{Christodoulou:2014yia, Yang:2018arj, Zhang:2019abv, Wang:2020fgz}& \qquad $\frac{3}{2}m$  & 1.5 & $3 \pi v \sqrt{3}m^2$ \\  
  2. & RN \citep{Han:2018jnf, Wang:2018txl, Ali:2018sqk}&\qquad $\frac{1}{4}(m^2 +\sqrt{9m^2-8q^2})$ &$1.4$& $\frac{\pi v \sqrt{(\sqrt{9m^2-8q^2}+3m)^2(m\sqrt{9m^2-8q^2}+3m-4q^2)}}{2\sqrt{2}}$ \\ 
3. & Kerr \citep{Wang:2018dvo, Wang:2019ake, Wang:2019dpk}& To find the maximal space-like hyper-surface, an induced metric is required on the space-like hyper-surface in the interior of the Kerr BH in such a way that  $r$ coordinate takes its largest value, also see appendix (\ref{A}).& $1.402$ &$2 \pi v \sqrt{-\Delta } \left(\sqrt{a^2+r^2}+\frac{r^2 \ln  \left(\sqrt{a^2+r^2}+a\right)}{(2 a) \left(\sqrt{a^2+r^2}-a\right)}\right)$ \\
4. & Kerr Newman \citep{Wang:2019ear, Ali:2021kdu}& Need the same procedure as that of Kerr BH&1.3005 &  $2 \pi v \sqrt{-\Delta' } \left(\sqrt{a^2+r^2}+\frac{r^2}{2 a}\ln \left(\frac{\sqrt{a^2+r^2}+a}{\sqrt{a^2+r^2}-a}\right) \right)
$ \\
5. & BTZ \citep{Zhang:2019pzd, Ali:2020olc}& \qquad $\sqrt{\frac{m}{2}}$ & $0.71$ & \qquad $\frac{\pi v}{2} m$ \\
6. & Rotating BTZ \citep{Ali:2020qkb, Maurya:2022vjd}& \qquad $\sqrt{\frac{l^2 m (\sqrt{3X^2+1}+2)}{6}}$ &$ 0.45$ & $\frac{\pi v}{3}(\sqrt{l^2 m^2 (-3X^2+2\sqrt{3X^2+1} +7)-9 J^2}$\\
7. & Charged  f(R) \citep{Ali:2019icq, Wen:2020thi}& \quad $\sqrt{6} \left(\sqrt{\frac{G}{R_0}-\frac{18 \sqrt{6} m}{b R_0 \sqrt{\frac{F}{R_0}}}}+\sqrt{\frac{F}{R_0}}\right)$ &- - - - - - -&$\frac{2}{3}\pi v \sqrt{\frac{2\sqrt{6} m}{b}A^3-6 q^2 A^2+\frac{A^6}{72 R_0}+A}$ \\
8. &  Neutral toral BH \citep{Ong:2015tua} & \qquad $\left (\frac{ m L^2}{\pi K^2}  \right )^{\frac{1}{3}}$&depends on the values of $n$, $K$ \& $L$& $\frac{8\pi m Lv}{n}$\\
 \hline
  \hline
\end{tabular}
\end{table}
here, 

$$G=\frac{4 b \left(3 q^2 R_0-4\right)}{\sqrt[2]{Z}}-\frac{\sqrt[3]{Z}}{b}+16, \qquad F=-\frac{4 b \left(3 q^2 R_0-4\right)}{\sqrt[3]{Z}}+\frac{\sqrt[3]{Z}}{b}+8, \qquad  A=\sqrt{\frac{F}{R_0}}+\sqrt{\frac{Y}{R_0}},$$

$$Y=-\frac{18 m \sqrt{6 R_0}}{b \sqrt{-4 b \sqrt[-3]{Z} \left(3 q^2 R_0-4\right)+\frac{\sqrt[3]{Z}}{b}+8}}+4 b \sqrt[-3]{Z} \left(3 q^2 R_0-4\right)-\frac{\sqrt[3]{Z}}{b}+16,$$
$K$ is the compactification parameter and $L^2$ is the AdS length scale.
\end{landscape}

\begin{equation}
\frac{1}{\sqrt{-g}}{{\partial_ \mu }\left(\sqrt{-g} g^{\mu v } \partial_v \Phi \right)}=0,
\end{equation}
with $\Phi$ as the scalar field for a respective BH, we get

\begin{equation}\label{Eqmot1}
    P^{\mu}P_{v}=g^{\mu v}P_{\mu}P_{v}=g^{00}E^2+h^{ij}P_iP_j, 
\end{equation}
$h^{ij} \quad (i, j =1,2, 3)$ is the $4-$dimensional induced inverse metric on the hyper-surface at a constant radius. It may be diagonal or non-diagonal but due to its quadratic property, it can be diagonalized by the coordinate transformation. So, we can get from Eq. (\ref{Eqmot1} )as 
\begin{equation}
E^2-\lambda^{ij}P_iP_j=0, \qquad \Rightarrow \qquad P_1^2=\frac{1}{\lambda^1}(E^2-\lambda^2P^2_2 -\lambda^3P^2_3),
\end{equation}

with $\lambda^{ij}$ as the diagonal elements (see appendix (\ref{B})). In $4d-$dimensional space-time the total number of quantum states are 
\begin{equation*}\label{Qmode}
    \begin{aligned}
        g(E)=\frac{1}{(2\pi)^3}\int dx_1 dx_2 dx_3 dP_1 dP_2 dP_3\\ =\frac{1}{(2\pi)^3}\int dx_1 dx_2 dx_3 dP_2 dP_3 \sqrt{\frac{1}{\lambda^1}}\sqrt{E^2-\lambda^2P^2_2 -\lambda^3P^2_3},
    \end{aligned}
\end{equation*}
\begin{equation}
    =\frac{E^3}{12\pi}V_{\sum},
\end{equation}
In the case of lower dimensions, one can determine the number of quantum states as
\begin{equation*}
    g(E)=\frac{1}{(2\pi)^2}\int dx_1 dx_2 dP_2\sqrt{\frac{1}{\lambda^1}}\sqrt{E^2-\lambda^2P^2_2},
\end{equation*}
\begin{equation}\label{Qmodes}
g(E)=\frac{E^2}{8\pi}V_{\sum},
\end{equation}
here the integral formulae $\int\int{\sqrt{1-\frac{x^2}{a^2}-\frac{y^2}{b^2}}}dx=\frac{2\pi}{3}ab, \quad \& \quad \int_{0}^{a}{\sqrt{1-\frac{x^2}{a^2}}}dx=\frac{\pi}{4}a \quad$ are used for calculating the total number of quantum states. Next, defining the free energy $F(\beta)$ for an inverse temperature $\beta=\frac{1}{T}$ is 
\begin{equation}\label{FE}
F(\beta)=\frac{1}{\beta}\int {ln(1-exp(-\beta E))}dg(E)=-\int{g(E)\frac{dE}{e^{\beta E}-1}},
\end{equation}
So, using the value of total quantum states one can easily determine the free energy for either $4-$dimension or $3-$dimensions space times. Hence, the quantum mode entropy of the massless scalar field is 
\begin{equation}\label{Ent1}
S_{\sum}=\beta ^2\frac{ \partial F(\beta)}{ \partial\beta}=\frac{\varpi}{\beta^2}\sqrt{-r_v ^2f(r_v)}v,
\end{equation}
here, $\varpi$ is some constant.  From this Eq. (\ref{Ent1}), the quantum mode entropy is also linearly proportional to $v$ i.e., any change in the quantum mode entropy could lead us to the change of the statistical quantities in the interior of the BH. 

We can calculate the variation of the statistical quantities in the interior BH using the relation of quantum mode entropy and advance time. For this, consider two assumptions that could lead us to find the evaluation relation between BH's interior and exterior entropy. These assumptions are

\begin{itemize}
\item \textbf{BHs radiation as black body radiations:} This assumption guarantees the BH temperature as seen by an observer from infinity is the same as event horizon temperature so, the Boltzmann law can be used for investigating the emission of radiation process \citep{landsberg:1989}. In $(n+1)$ dimension space-time of a symmetric BH, the Boltzmann law is 
\begin{equation}\label{dv}
\frac{dm}{dv}=-\sigma A T^{n+1} \Rightarrow dv=-\frac{\beta^{n+1} \gamma }{ A }dm,
\end{equation}
here, $n$ is the number of space dimensions, and $A$ is  the event horizon area of the BH.
\item \textbf{Quasi-static process:} The radiation emission is a quasi-static process. It means that the evaporation process is much slow i.e. $\frac{dm}{dv}<<1$ but Hawking's temperature continuously changes i.e., the thermal equilibrium between the scalar field and horizon of the BH is adiabatically preserved $(\Delta Q=0)$. This assumption guarantees the investigations of the variation of BH's radiation for an infinitely small  time interval ( quantum level.)
\end{itemize}
With these two assumptions, we can write the differential form of quantum mode entropy from Eq. (\ref{Ent1}) as
$$ \dot{S}_{CR}=\frac{\pi^2 \dot{V}_{CR}}{45\beta^{n+1}},$$
for a $4-$dimensional system $n=3$, so
\begin{equation}\label{diffEnt1}
    \dot{S}_{CR}=\frac{\pi^2 \dot{V}_{CR}}{45\beta^4},
\end{equation}
here, we need to find the differential volume for the respective BH and fit in this Eq. (\ref{diffEnt1})  to get the differential form of quantum mode entropy. As the interior volume of a BH is a function of advance-time so, one can easily take the differentiation of interior volume concerning advance-time and fit Eq. (\ref{dv}) in it to get the differential volume in terms of $\rm A, ~ \beta, ~\& ~M, ~Q, ~a$ (depending on the nature of BH geometry). To get the evaporation relation between the interior and exterior entropy of the BH, we need to compare the quantum mode entropy with Hawking entropy. It can be done in two ways: 
\begin{itemize}
    \item directly find the differential form of Hawking entropy for the respective BH i.e. 
\begin{equation}\label{diff BH}
    \dot{S}_{BH}=\frac{\dot{A}}{4\pi},
\end{equation} 
or
    \item use the first law of BH thermodynamics. For spherically symmetric rotating BHs the $1^{st}$ of BH thermodynamics is
\begin{equation}\label{1stlaw}
{dm}=\frac{d{S}_{BH}}{\beta},
\end{equation}
here $S_{BH}$ is Hawking's entropy of BH. In the case of rotating BHs, a spherically symmetric BH is considered thus, we can ignore exotic features of charge and angular momentum.
\end{itemize}

\begin{landscape}
\begin{table}[!th]
\caption{The interior entropy for different BHs}
\label{tab:2}
\begin{tabular}{||c|p{3.5cm}|p{8.5cm}|p{6cm} || }
\hline
 \hline
  S. No. &Black hole  
&Entropy $(S_{\sum})$&Relation b/w the variation of Interior Exterior Entropy  $\left(\frac{\dot{S}_{\sum}}{\dot{S}_{BH}}\right)$\\ 
 \hline
 \hline
1. & Schawarzschild \citep{Christodoulou:2014yia, Yang:2018arj, Zhang:2019abv, Wang:2020fgz}
& $\frac{3\sqrt{3}\gamma}{(90\times 8^4)\pi} A$ & $-\frac{\sqrt{3}\pi^2 \gamma}{240}$ \\  
2. & RN\citep{Han:2018jnf, Wang:2018txl, Ali:2018sqk}
& $\frac{v \left(m^2-q^2\right)^{3/2} \sqrt{m \left(\sqrt{9 m^2-8 q^2}+3 m\right)^2 \sqrt{\sqrt{9 m^2-8 q^2}+3 m-4 q^2}}}{720 \sqrt{2} \left(\sqrt{m^2-q^2}+m\right)^6}$ &$-\frac{\pi^2 \gamma}{360 \sqrt{2}}F(m,q)$  \\
3. & Kerr  \citep{Wang:2018dvo, Wang:2019ake, Wang:2019dpk}&$\frac{\pi ^2  \left(m^2-a^2\right)^{3/2}}{360 \pi \left(\left\{\left(\sqrt{m^2-a^2}+m\right)^2+a^2\right\}\right)^3}V_{\sum}$ &$-\frac{\pi^2\gamma}{180}F\left(\frac{a(v)}{m(v)}\right)$  \\
4. &  Kerr Newman \citep{Wang:2019ear, Ali:2021kdu}&$\frac{1}{180}\frac{f_{max}(m,a,q)(m^2-a^2-q^2)^{\frac{3}{2}}}{(2m\sqrt{m^2-a^2-q^2}+2m^2-q^2)^3}v$ &$-\frac{\pi^2 \gamma}{90}F(m,a,q)$ \\
5. &BTZ \citep{Zhang:2019pzd, Ali:2020olc}
&$\frac{3\zeta(3)}{2\beta^2}\sqrt{-r_v^2 f(r_v)}$ &$-\frac{6\zeta(3) \gamma}{\pi}\frac{\sqrt{-r_v^2 f(r_v)}}{r_+}, \quad r_+=\sqrt{m}$ \\
6. & Rotating BTZ \citep{Ali:2020qkb, Maurya:2022vjd}
&$\frac{3\zeta(3)}{2\beta^2}\sqrt{-r_v^2 f(r_v)}v$ &$-\frac{4\gamma\zeta(3)}{3\pi}\frac{\sqrt{m(3X^2+\sqrt{3X^2+1}-1)}}{\sqrt{X+1}}$  \\
7. & Charged  f(R) \citep{Ali:2019icq, Wen:2020thi}
&$\frac{\pi^2 V_{\sum}}{45 \beta^3}=\frac{1}{4320}\alpha(m,q; b)$ & $-\frac{4\pi^2\gamma}{135}\gamma'(m,q;b)$ \\
8. & Neutral toral BH \citep{Ong:2015tua} &$-\frac{4 l m^2 v}{45 \pi  K^2 L^4}$ &$\frac{\gamma }{30}  \sqrt[3]{\frac{\pi ^{11} K^{10} m}{L}}$ \\
 \hline
  \hline
\end{tabular}
\end{table}

Here, we used $V_{\sum}$ as the volume of a respective BH as given in Table (\ref{tab:1}). The dot $(.)$ is used for differentiation concerning Eddington time $(v)$, $S_{BH}$ is the Bekenstein and Hawking entropy of the respective BH that is equal to one-quarter of its area. 
\begin{equation*}
\begin{aligned}
    F(m, q)=\frac{\sqrt{(\sqrt{9m^2-8q^2}+3m)^2{(m\sqrt{9m^2-8q^2}+3m-4q^2)}}}{\sqrt{m^2-q^2}+m},\qquad F\left(\frac{a_o[v]}{m_o[v]}\right)=f_{max}\left(\frac{a_o[v]}{m_o[v]}\right)\left[1-\sqrt{1-\left(\frac{a_o[v]}{m_o[v]}\right)^2}\right]\left(\frac{a_o[v]}{m_o[v]}\right)^{-2},\\ F(m, a,q)=\frac{f_{max}(m,a,q)(m^2-a^2-q^2)^{\frac{3}{2}}}{(2m\sqrt{m^2-a^2-q^2}+2m^2-q^2)^3}, \qquad f_{max}(m,a,q)=\sqrt{2mr-r^2-a^2-q^2}\left(\sqrt{r^2+a^2}+\frac{r^2}{2a}\frac{\sqrt{r^2+a^2}+a}{\sqrt{r^2+a^2}-a}\right)_{r=r_v},
\end{aligned}
\end{equation*}

\end{landscape}

Now, either dividing Eq.  (\ref{diffEnt1}) by Eq. (\ref{diff BH}) or fitting Eq. (\ref{1stlaw}) in Eq. (\ref{diffEnt1}) one gets the evaluation relation of the two types of entropy. Some results are summarized in table (\ref{tab:2}). The negative sign in proportional relation represents the increase of quantum modes entropy with advance or Eddington time whereas, Hawking's entropy decreases. Using these results, the effect of mass, charge, and angular momentum can also be elaborated from the curves (for detailed insight, see also the references given in the table (\ref{tab:2}));  

These investigations attract many authors due to their universality in explaining the BH's interiors with the Einstein theory of general relativity. The dynamical nature of these investigations with $v$  could be used for BH's interior volume to probe the interior information of BHs.

The structure of this paper is such that in the section (\ref{terminology}), we present the meanings of important terms for the readers to understand the concept of the work done on the topic of the subject. The main literature on the BH interior volume, entropy, and its connection to the evaporation and information paradox is presented in section (\ref{introduction}). To explain work more simply, we present the review of interior volume, interior entropy, and their variation for Schawarzschild BH, Kerr BH, and BTZ BH to cover the $(3+1)$ as well as $(2+1)$ dimensional space-times in section (\ref{Schw-BH}), (\ref{Kerr-BH}), and (\ref{BTZ-BH}). In section (\ref{ent-variation}), the entropy variation and evaporation relation for different BHs are being explained. In section (\ref{natur-radiation}), we in-sighted the probability of emission and the nature of BH radiation. Finally, some remarks and discussion on the main point of the results are presented.
\section{A review of the BH interior Volume and Entropy variation}\label{vol-ent-review}

\subsection{Schwarzschild BH}\label{Schw-BH}

In Eddington Finkelstein coordinates $(~v,~r,~\theta,~\phi)$ the line element of Schwarzschild geometry is given in Eq. (\ref{collapsedmass}) e.g., see Refs. \citep{Christodoulou:2014yia, Zhang:2015, Wang:2018dvo, Ali:2021kdu}. For an uncharged spherically symmetric BH $f(r)=1-\frac{2m}{r}$ and the $2$-sphere form is $d\Omega^2=r^2(d\theta^2+sin^2 \theta d\phi)$ and the advance time is defined as $v=t+r^*$.
\begin{equation}
r^*=r+2mlog|r-2m|,
\end{equation}

here, the geometric units $G=c=\hbar=k_B=1$. The hyper-surface on the proposed sphere could be defined as the product of an affine parameter to the 2-sphere i.e., $\sum=\gamma\times S^2$,  where $\gamma\rightarrow(v(\lambda),r(\lambda))$ is the affine parameter. So, the Schawarzschild metric becomes
\begin{equation}\label{EDmet}
{ds}^2_{\sum}=\left(-f(r) \dot{v }^2+2 \dot{v } \dot{r}\right)d \lambda ^2+r^2d\Omega^2,
\end{equation}

We have considered only curved space-time so, the contribution to the volume will not be the same as $V=\frac{4}{3} \pi R^3$. Using Eq. (\ref{EDmet}) the interior volume of Schawarzschild BH is
$$V_{CR}=\int  \int d\lambda d\Omega \sqrt{r^4\left(-f(r)\dot{v }^2+2 \dot{v } \dot{r}\right)sin^2 \theta},$$
\begin{equation}\label{intvolgenfor}
=4 \pi\int{\text{d$\lambda $}}\sqrt{r^4\left(-f(r)\dot{v }^2+2 \dot{v } \dot{r}\right)},
\end{equation}

It shows that the proper length of a geodesic in auxiliary metric ($4\pi$-times) is the volume bounded by $\sum$. By maximizing this auxiliary metric, we need $\dot{r}=0$. So, we can find numerically the largest curve bounding the  maximal interior volume at $r=\frac{3}{2}M$. This equation also shows that finding the hyper-surface is similar to solving the geodesic equation to get the equation of motion (EOM) with Lagrangian. So, the auxiliary metric on the hyper-surface becomes; 
\begin{equation}\label{effmetric}
{ds}^2_{eff}=r^4 \left(-f (r) \dot{v }^2+2 \dot{v } \dot{r}\right),
\end{equation}
This equation shows that calculating the volume of the hyper-surface $V_{\sum}$ is equivalent to calculating the volume from auxiliary metric with Lagrangian whose maximal value is

$$L\left(r,v, \dot{r}, \dot{v}\right)=1,$$
i.e.
\begin{equation}\label{Eq.1}
r^4\left(-f(r) \dot{v }^2+2 \dot{v } \dot{r}\right)=1,
\end{equation}
using this Eq. (\ref{intvolgenfor}), we gets 
\begin{equation}\label{VCR}
V_{CR}=4\pi \lambda_f,
\end{equation}

As the metric $\tilde{g}_{\alpha\beta}$   has a Killing vector i.e., $\zeta^\mu=(\partial_\mu)^\mu\propto (1, 0) $ and $\gamma$ is an affinely parameterized geodesic in auxiliary metric so, the inner product of Killing vector $\zeta^\mu$ with its tangent $\dot{x}^\alpha=(\dot{v}, \dot{r})$ will be conserved

\begin{equation}\label{Eq.2}
\zeta\times \dot{x}^\alpha=r^4\left(-f(r) \dot{v }^2+2 \dot{r}\right)=Y,
\end{equation}

solving Eq. (\ref{Eq.1}) and (\ref{Eq.2}), we get
\begin{equation}\label{Eq.3}
\dot{r}=-r^{-4}\sqrt{A^2+r^4f(r)}, \qquad and  \qquad \dot{v}=\frac{1}{Y+r^4\dot{r}},
\end{equation}

In the case of space-like geometry, $(-f(r) \dot{v }^2+2 \dot{v } \dot{r})>0$ so the Lagrangian $L>0$. since $r$ is positive and the Lagrangian will demolish at $r=0$, which is the endpoint of the geodesic. Hence from Eq. (\ref{Eq.3}), $\dot{r}$ becomes infinite. Thus, $\gamma$ is a space-like geodesic in $m_{aux}$. A well-suited parameterization is to take $\lambda$ as the proper length in auxiliary metric. So, the auxiliary metric given above becomes,
\begin{equation}\label{auxmet}
ds_{M_{aux}}=-\sqrt{-r^4f(r)}dv=Ydv,
\end{equation}

It can be easily seen that $Y$ has to be negative for the geodesic to be space-like. Then, $\dot{v}$ and $\dot{r}$ are both negative and there are only positive terms in (\ref{Eq.1}). Integrating (\ref{Eq.2}), we get
\begin{equation}\label{VCR/4pi}
\frac{V_{\sum }}{4 \pi }=\lambda _f=\int _0 ^{2 M}\frac{r^4}{\sqrt{Y^2+r^4f(r)}},
\end{equation}
Eq. (\ref{VCR/4pi}) shows the restriction could be imposed on $Y$ as
$$Y^2> -r_v^4f(r_v)>=0,$$
\begin{equation}
\Rightarrow Y^2=\frac{27}{16}M^4=Y_c ^2,
\end{equation}

This condition is obtained from the expression $(-f(r) \dot{v }^2+2 \dot{v } \dot{r})>0$ having the roots $r=0$ and $r=2m$ otherwise, the position is maximum at $r_v=\frac{3}{2}m$. Since $(-f(r) \dot{v }^2+2 \dot{v } \dot{r})>0$ in the range $0<r<2m$. For every constant $r$, there is a solution or we can say $r$ is constant the surface is space-like geodesic of an auxiliary manifold. For a stationary (maximal) point of the volume given in Eq. (\ref{Eq.3}(b))
$$\frac{dv}{d\lambda}=\frac{1}{Y}\Rightarrow d\lambda=Hdv,$$
Integrating, we get
\begin{equation}
\lambda_f=H(v_f-v),
\end{equation}
As $r =\rm constant$, the surface will have the maximal volume between two given values of advance time $v$ when $H$ is largest. This means that, $r=r_v$ which gives $H=H_c$. So, the volume will be the largest possible. These considerations provide the basis for the derivation of the asymptotic volume.
\begin{equation}\label{Acnu}
\lambda_f=H_c v,
\end{equation}
Using Eq. (\ref{Acnu}) in Eq. (\ref{VCR}), one could get the interior volume as
\begin{equation} \label{VCR1}
V_{CR}=-4\pi\sqrt{-r^4f(r)}v=4\pi H_c v,
\end{equation}
or we can write as
\begin{equation}\label{SCHVCR}
V_{CR}=3\sqrt{3}\pi m^2 v,
\end{equation}

Where $A=-\sqrt{-r^4f(r)}$, Which shows that the interior volume depends on advance time. This result can be extended to other cases by simply using metric Eq. (\ref{EDmet}) with lapse function $f(r)$ from the desired BH metric and finding the maximization of $H=H_c$, to get the asymptotic expression, using an analogy to the above equation, one gets the required BH interior volume.  Calculating the interior volume of charged and charged f(R) BH from Eq. (\ref{VCR1}). is convenient. For this, one needs to calculate the factor $\sqrt{-r^4f(r)}$ for $r=r_v$ from the lapse function of the respective BH metric as given in (\ref{tab:1}).

\begin{figure}
\begin{center}
\includegraphics[width=0.45\textwidth]{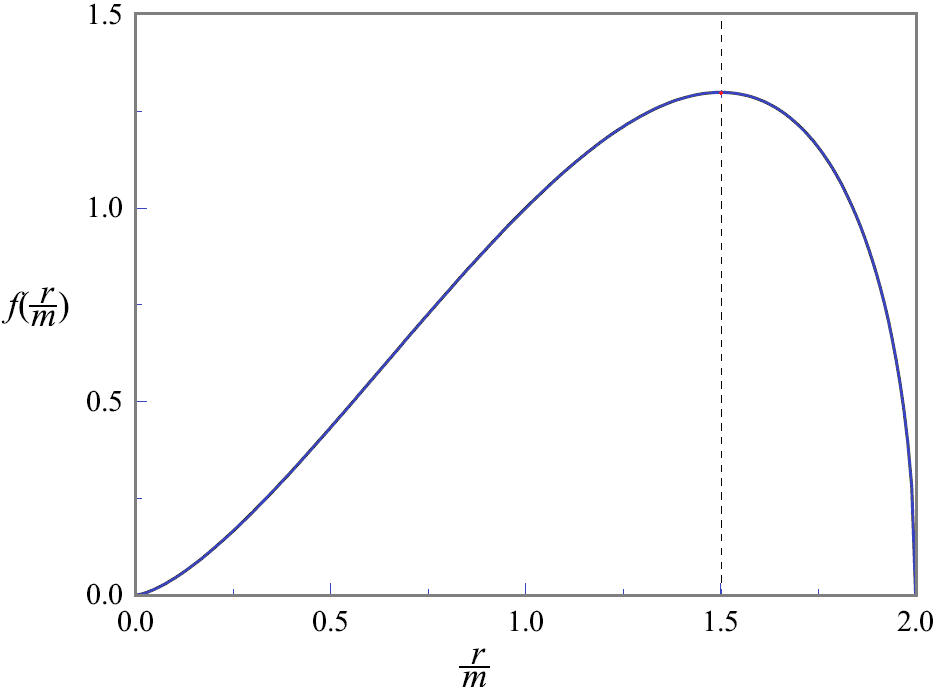}
\caption{The plot of proportionality function $f_{max}\left(\frac{r}{M}\right)$ vs $\frac{r}{M}$ for Schawarzschild BH showing the position of the maximal hyper-surface at $(\frac{r}{M})=\frac{3}{2}$ \citep{Wang:2018dvo}.}
\label{image-1.4}
\end{center}
\end{figure}

In Ref. \citep{Wang:2018dvo}, used the $a=0$ to degenerate the result of the Schawarzschild BH from the Kerr BH result and found the proportionality function for maximal hyper-surface as \citep{Wang:2018dvo}
\begin{equation}
    f\left(\frac{r}{m}, \frac{a}{m}\right)=f\left(\frac{r}{m}\right)_{max}=\frac{r}{m}\sqrt{2\frac{r}{m}-\left(\frac{r}{m}\right)^2},
\end{equation}
and the position of the maximal hyper-surface is plotted in Fig. (\ref{image-1.4}).

Baocheng Zhang \citep{Zhang:2015} considered a scalar field in the interior of Schwarzschild's BH and discussed its quantum modes entropy by determining the quantum states in the interior volume labeled by $(\lambda, \theta, \phi, p_{\lambda}, p_{\theta}, p_{\phi})$. So, the quantum mode corresponding to the cell of volume $(2\pi)^3$ in that phase space will have the total number of quantum modes given by
\begin{equation}
\frac{phase{\quad}space}{volume{\quad}of{\quad}the{\quad} cell}=\frac{d\lambda d\theta d\phi dp_{\lambda}dp_{\theta} dp_{\phi}}{(2\pi)^3},\qquad \hbar=1,
\end{equation}
For the total quantum modes, one needs to integrate the above expression as
\begin{equation}\label{freeEnergy}
g(E)=\frac{1}{(2 \pi )^3}\int{d\lambda d\theta d\phi dp_{\lambda}dp_{\theta} dp_{\phi}},
\end{equation}
This  gives the total number of quantum modes having energy less than $E$. As from the WKB approximation, the scalar field in the interior of a BH consisting of quantum modes is
\begin{equation}\label{Scalarfield}
\Phi =e^{\text{-iET}}e^{\text{iI}(\lambda, \theta, \phi)},
\end{equation}
Expanding and solving the above equation in components $(\lambda, \theta, \phi)$ and solving, gives the EOM (e.g., see: \ref{A})
\begin{equation}\label{eqofenergy}
E^2-\frac{p_{\lambda }^2}{-f(r) \dot{v }^2 +2 \dot{v } \dot{r}}-\frac{p_{\theta }^2}{r^2}-\frac{p_{\phi }}{ r^2 \sin ^2 \theta }=0,
\end{equation}
here, we used
$$\frac{\partial I}{\partial \lambda }=p_{\lambda }, \qquad \frac{\partial I}{\partial \theta }=p_{\theta }, \qquad \frac{\partial I}{\partial \phi }=p_{\phi },$$
as eigenstates of the diagonal elements of the  metric. Solving Eq. (\ref{eqofenergy}) for $p_\lambda$, we get
\begin{equation}\label{P:lambda}
p_{\lambda }=\sqrt{-\dot{v }^2 f(r)+2 \dot{v } \dot{r}} \sqrt{E^2-\frac{p_{\theta }^2}{r^2}-\frac{p_{\phi }}{r^2 \sin ^2\theta}},
\end{equation}
Use Eq. (\ref{P:lambda}) in Eq. (\ref{freeEnergy}) and simplifying, we get 

\begin{equation}
    \begin{aligned}
         g(E) \\ =\int{d\theta  d\lambda  d\phi\sqrt{- f(r)\dot{v }^2+2 \dot{v } \dot{r}} \int{\sqrt{E^2-\frac{p_{\theta }^2}{r^2}-\frac{p_{\phi }}{r^2 \sin ^2\theta}}}d {\theta } dp_{\phi }},
    \end{aligned}
\end{equation}

$$=\frac{1}{(2 \pi )^3}\int d\theta  d\lambda  d\phi\sqrt{- f(r)\dot{v }^2+2 \dot{v } \dot{r}} \left (\frac{2 \pi }{3} E^2 r^2 sin^2 \theta  \right ),$$
$$=\frac{E^3}{(2 \pi )^3}\frac{2 \pi }{3}\int{d\lambda  \sqrt{r^4 \left(2 \dot{v } \dot{r}-\dot{v }^2 f(r)\right)}}{\int{\sin^2 \theta d\theta} \int{d\phi}},$$
$$=\frac{E^3}{12 \pi ^2}\left ( 4\pi\int{d\lambda  \sqrt{r^4 \left(2 \dot{v } \dot{r}-\dot{v }^2 f(r)\right)}} \right ),$$
\begin{equation}\label{Qstate}
g(E)={\frac{E^3}{12 \pi ^2}{V_{CR}}},
\end{equation}

For the integration in the first step, we used the general formula $ \sqrt{1-\frac{x^2}{a^2}+\frac{y^2}{b^2}} dx dy=\frac{2 \pi }{3} a b$. As from Eq. (\ref{Qstate}),  $g(E)\propto V_{CR}$ so, it still has similarity with normal space-time. Note that the physical interpretation does not need to be the same because the volume in general relativity results from the curved space-time. The main reason for the physical difference can be considered as the volume is bound within the closed hyper-surface that is increasing with advanced time $v$. So, the number of quantum states inside the BH must also increase with time. This statement is crucial for interior volume and information storage in curved space-time. Considering this difference, we can calculate the free energy as
\begin{equation}\label{BfreeE}
F(\beta )=\frac{1}{\beta }\int {dg(E)} ln(1-e^{-\beta (E)}),  
\end{equation}
$$ F(\beta)=-\int{\frac{dg(E)}{-1+e^{-\beta(E)}}},$$
$$F(\beta)=-\frac{V_{\text{CR}}}{12 \pi ^2}\int{\frac{E^2 dE}{-1+e^{-\beta (E)}}},$$
Solving the integral, we get the result as
\begin{equation}\label{FreeEnergy}
F(\beta)=-\frac{\pi^2 V_{CR}}{180\beta^4},
\end{equation}
Finally, entropy is
\begin{equation}\label{Ent11}
S_{CR}=\beta^2 \frac{\partial F}{\partial \beta}=\frac{\pi^2 V_{CR}}{45\beta^3},
\end{equation}

Which is the entropy in the interior of the $V_{CR}$. Using the value of interior volume $V_{CR}$ from Eq. (\ref{SCHVCR}) and inverse temperature, we can get the quantum modes entropy in BH's interior. Since a BH has the property of emitting radiation that is claimed to be quasi-static and increases with time due to variable temperature. Treating BH radiation as black body radiation can be defined by Stefan Boltzmann's law. Thus, the rate of mass loss from the Schwarzschild BH due to Hawking's radiation is
\begin{equation}\label{Boltzmannlaw}
\frac{dM}{dv}=-\frac{1}{\gamma M^2},
\end{equation}
This equation states that the time duration for the radiation to last from a BH is proportional to the triple power of mass $M$ i.e.,
$$v\approx\gamma M^3,$$
This also satisfies the condition of Ref. \citep{Christodoulou:2014yia} with $v>>M$.
Now, consider Schwarzschild BH inverse temperature as $\beta=\frac{1}{T}= 8\pi M$, then the interior entropy is 
\begin{equation}
S_{CR}=\frac{3 \sqrt{3} \gamma  M^2}{45\times 8^3}=\frac{3 \sqrt{3}\gamma A}{(45\times 8^4) \pi },
\end{equation}
here $ A=16\pi M^2$ is Schwarzschild BH's surface area  \citep{Bekenstein:1972tm, Zhang:2015}, as the radiations from the BH are in the Planks scale and the final evaporation stage has yet not been discovered, this means that the mass loss of the BH during the emission of radiation is so small. Hence, we can take $\frac{dM}{dv}\approx M$, which is inconsistent with the requirements of the CR volume because in such a case we can't take the growth in the BH's volume.  The above equation confirms  that the entropy of the quantum field in the CR volume is directly related to the horizon area and also the coefficient of $A$ is much smaller than $\frac{1}{4}$. This means it doesn't satisfy the first law of BH satisfied by the Bekenstein and Hawking relation. From this relation, we can say there is more information loss on the BH horizon. Here we can also raise the question of how to fit the above relation in the first law of BH thermodynamics. Let's compare the exterior and interior entropy that may have some entangled relation. This may justify these issues with the $1^{st}$ law of BH's thermodynamics.  So, considering the quasi-static emission of radiation, we can introduce the differential form of Eq. (\ref{Ent11}), and further, it can be represented in Hawking's entropy as

\begin{equation}\label{difEnt}
dS_{CR}=\frac{\pi^2 dV_{CR}}{45\beta^3}=-\frac{\sqrt{3}\pi^2\gamma}{30}mdm= -\frac{\sqrt{3}\pi\gamma}{240}(S_{BH}),
\end{equation}
This shows that interior quantum mode entropy is directly related to Hawking entropy. In this equation, the negative sign shows that as the Horizon entropy increases, the interior entropy decreases due to the loss of BH's information.

\subsection{Kerr BH}\label{Kerr-BH}
In the case of  Kerr BH \citep{Kerr:2007dk}, the interior and exterior horizons can be calculated from $\Delta=0$,
\begin{equation}\label{Kerrrad}
r_\pm=M\pm \sqrt{M^2-a^2},
\end{equation}
Due to the spinning property, the horizons of a Kerr BH may not be spherically symmetric and this character may affect the physical properties of Kerr's BH. As discussed in the Penrose diagram of CR's work, the interior volume is mainly the contribution of  the stretched part of the largest spherically symmetric $3d$ hyper-surface that can be calculated by either using the condition of vanishing curvature or by maximization of the factor $\sqrt{-r_v^4 f(r_v)}$. We also stated that this part of the hyper-surface doesn't extend to singularity, so one can say that the spinning character doesn't affect the interior volume of a Kerr BH. Without loss of generality, for a Kerr BH, the  interior volume can be defined as
\begin{equation}
V_{CR}=\int{\sqrt{-\Delta}\rho sin\theta dv d\theta d\phi},
\end{equation}
that gives
\begin{equation}\label{KerrBHvol}
V_{\text{CR}}=2 \sqrt{-\Delta } \pi v \left(\sqrt{r^2 + a^2} + \frac{r^2}{2 a} log\left(\frac{\sqrt{a^2+r^2}+a}{\sqrt{a^2+r^2}-a}\right)\right),
\end{equation}

This result is also investigated in \citep{Bengtsson:2015zda}. At $r=r_s$, this equation gives the largest hyper-surface to calculate the maximal interior volume. Generally, this Eq. (\ref{KerrBHvol}) can be written as
\begin{equation}\label{Kerrvol}
V_{CR}=2\pi F(r,a)v =2\pi M^2 F_{max}\left(\frac{a}{M}\right)v,
\end{equation}
where 
$$F(r,a)=\sqrt{-\Delta } \left(\sqrt{r^2 + a^2} + \frac{r^2}{2 a} log\left(\frac{\sqrt{a^2+r^2}+a}{\sqrt{a^2+r^2}-a}\right)\right),$$
and 
$$F_{max}\left(\frac{r}{M}\right)=F\left(\frac{r_v}{M},\frac{a}{M}\right),$$

\begin{figure}[ht]
\begin{center}
\includegraphics[width=0.45\textwidth]{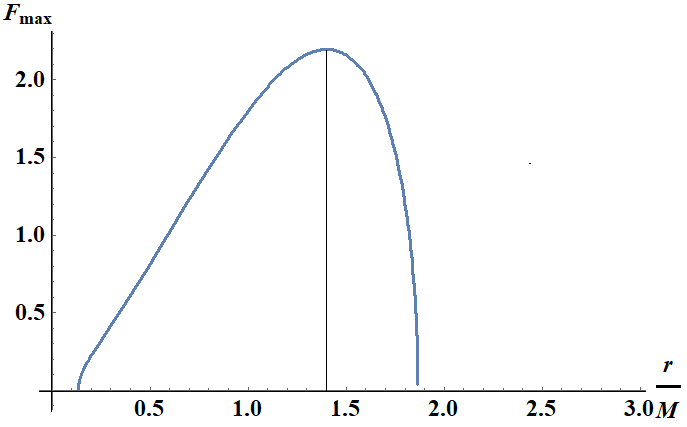}
\caption{The plot of $F_{max}\left(\frac{r}{M}\right)$ vs $\frac{r}{M}$ for a Kerr BH shows the proportional function has the  maximum value at $(\frac{r}{M})=1.402$ with $(\frac{a}{M})=0.2$}
\label{image-2.4}
\end{center}
\end{figure}

The position of the largest hyper-surface is numerically found at $r_v=1.402$ as shown in Fig. (\ref{image-2.4}).

This relation can degenerate the Schwarzschild result by using $a=0$. Plotting the proportional relation, we can obtain the same maximal hyper-surface at $r_v=1.5$ \citep{Wang:2018dvo}.

Following the Baocheng Zhang scenario \citep{Wang:2018dvo}, the entropy of static BHs (either charged or uncharged) is discussed in the above sections. We found that the entropy of the massless scalar field directly increases with $v$. So, this is the main character that affects the interior statistical quantities of the BH including its volume. To calculate the BH's entropy associated with quantum modes, we need to calculate the number of interior quantum states. Considering the metric for a maximal hyper-surface ( $r=r_v$), the volume is determined in Eq. (\ref{Kerrvol}). Next, considering the quantum modes of the massless scalar field in a Kerr BH, we obtain the EOM in general form as 
\begin{equation}
-E^2+h^{ab}P_a P_b=0,
\end{equation}

Where $h^{ab}$ is the auxiliary metric on the maximal hyper-surface. So, the total number of quantum states contained in the scalar field is given by Eq. (\ref{Scalarfield}). Where $V_{CR}$ is given in Eq. (\ref{KerrBHvol}). The free energy calculated is given in Eq. (\ref{FreeEnergy}) and the entropy in Eq. (\ref{Ent1}). Next, using two conditions. The BH emission rate is a quasi-static process (it needs the emission rate to satisfy the condition $v>>M$) and BH's radiation is black body radiation. The latter condition guarantees the horizon temperature as Hawking temperature given by
\begin{equation}\label{insvtemp}
\beta =\frac{1}{T}=\frac{4\pi \left(r_{\pm }+a^2\right)}{r_+-r_-},
\end{equation}

here, $r_\pm$ is given in Eq. (\ref{Kerrrad}), the first condition guarantees that the temperature of the scalar field and the horizon will be in equilibrium if the process is considered at the quantum level. So, the temperature of the scalar field will be considered equal to the temperature of the horizon, and the rate of mass loss can be seen by the Stefan Boltzmann law in Eq. (\ref{Boltzmannlaw}). Using the values of $\beta$ and $A$, we can get Eq. (\ref{Boltzmannlaw}) as
\begin{equation}\label{BoltzmannlawKerr}
\frac{dM}{dv }=\frac{\left(r_+-r_-\right){}^4}{32 \gamma \pi ^3 M^3 r_+^3}=\frac{\left(M^2-a^2\right)^2}{32 \gamma \pi ^3 M^3 \left(\sqrt{M^2-a^2}+M\right)^3},
\end{equation}
Where $\gamma$ is a positive constant, it depends on quantum modes coupled with gravity. Its value doesn't affect our discussion. As the volume of BH changes with advanced time hence, using the above equation, we can introduce the volume in terms of differential form Eq. (\ref{Kerrvol}) as
\begin{equation}\label{KerrvolBoltzmannlaw}
dV_{CR}=2 \pi M^2   F\left(\frac{r}{M},\frac{a}{M}\right)dv,
\end{equation}
In addition to Eq. (\ref{BoltzmannlawKerr}), and Eq. (\ref{KerrvolBoltzmannlaw}) gives
\begin{equation}
\dot{V}_{CR}=-64 \gamma \pi ^4 M^4 F_{\max }\left(\frac{r_{v }}{M},\frac{a}{M}\right)\frac{\left(\sqrt{M^2-a^2}+M\right)^3}{\left(M^2-a^2\right)^2}\dot{M},
\end{equation}
Finally, the differential entropy is
\begin{equation}
\begin{aligned}
    \dot{S}_{\text{CR}}=-\frac{\pi ^2}{45 \beta }\dot{V}_{CR} \\ =-\frac{8 \pi^3  \gamma  M^2}{45} \frac{ F_{\max }\left(\frac{r_{v }}{M},\frac{a}{M}\right)\left(\sqrt{M^2-a^2}+M\right)^3 \dot{M}}{\sqrt{M^2-a^2} \left(\sqrt{M^2-a^2}+a^2+M\right)^3} ,
\end{aligned}
\end{equation}

Now, to get the evolution relation between the exterior and interior entropy, we need the variation of Bekenstein and Hawking entropy for Kerr BH  \citep{Bekenstein:1972tm, Bekenstein:1973ur, Bardeen:1973gs}. As the Hawking entropy is
$$S_{BH}=\frac{A}{4}=\pi(r^2 _+ -a^2),$$
Where $A=4\pi (r_+^2+a^2 )$ is the area of the Kerr BH. Its differential form is 
\begin{equation}\label{DRNBHent}
\dot{S}_{BH}=\frac{2\pi(1+\sqrt{ M^2- a^2})^2}{\sqrt{ M^2- a^2}}M\dot{M},
\end{equation}
Here $\dot{M}<0$ so, the proportional relation between the two types of entropy can be written as 
\begin{equation}\label{KerrENT}
\dot{S}_{CR}=-\frac{\pi ^2}{180}\gamma f\left(\frac{r_{v }}{M},\frac{a}{M}\right)\dot{S}_{BH},
\end{equation}
here,
\begin{equation}
    \begin{aligned}
        f\left(\frac{r_{v }}{M},\frac{a}{M}\right)=f_{\max }\left(\frac{a}{M}\right)
     \\ =F_{\max }\left(\frac{r_{v }}{M},\frac{a}{M}\right)\left (1-\sqrt{1-\left(\frac{a}{M}\right)^2}  \right )\left ( \frac{a}{M} \right )^{-2}
\end{aligned}
\end{equation}

\begin{figure}
\begin{center}
\includegraphics[width=0.45\textwidth]{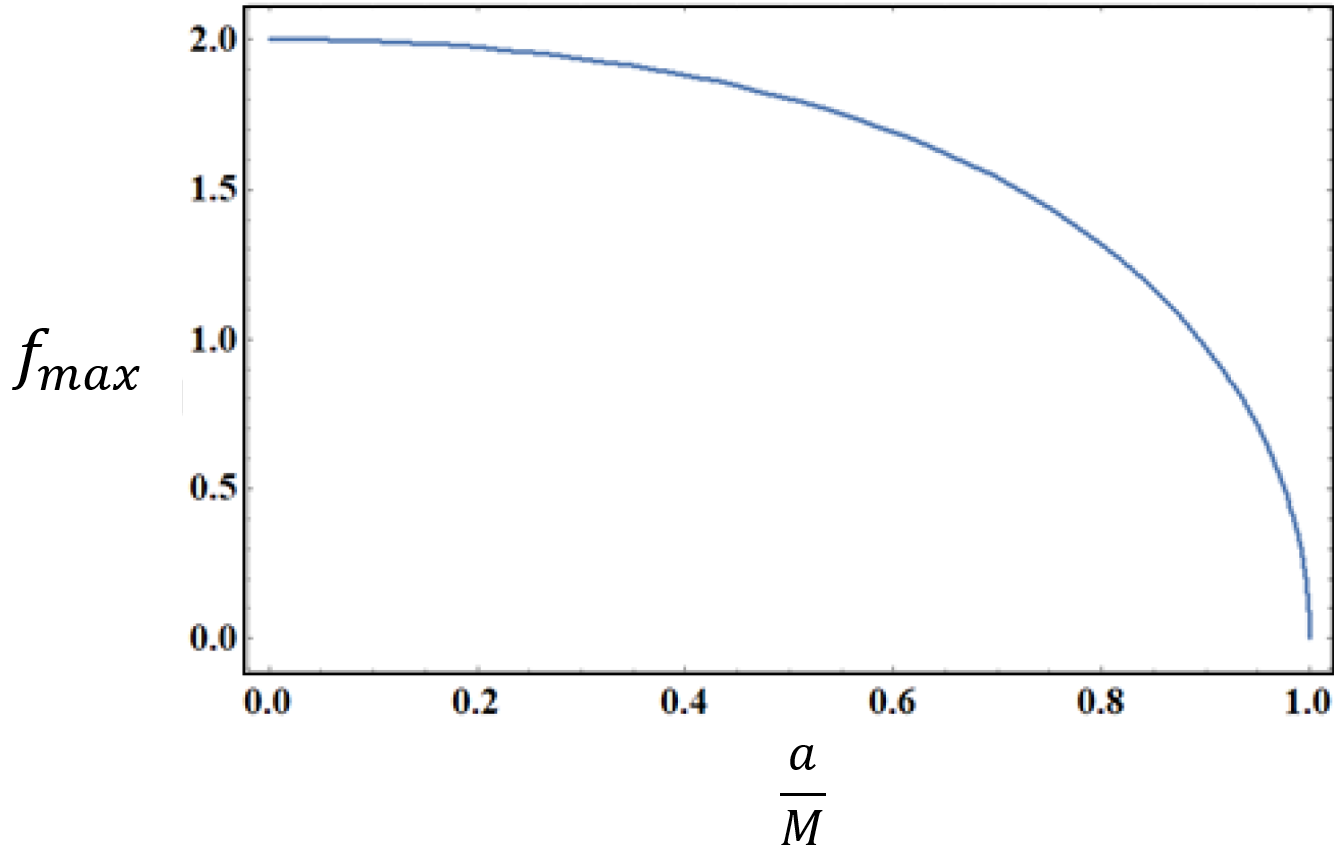}
\caption{The plot of $f_{max} (\frac{a}{M})$ vs $(\frac{a}{M})$ for a Kerr BH \citep{Wang:2018dvo}.}
\label{image-2.6}
\end{center}
\end{figure}

Alternatively, this result can be obtained by determining the volume bounded by the largest hyper-surface in the interior of Kerr BH and fitting its value in the general form of entropy Eq. (\ref{Ent11}). Next, using the two assumptions to determine the differential form and finally by comparison with Bekenstein and Hawking entropy, one gets the same result as Eq. (\ref{KerrENT}). The plot of this function $f_{max} (\frac{a}{M})$ vs $(\frac{a}{M})$ is given above. The proportionality function is plotted against the spin parameter in Fig. (\ref{image-2.6}). As the value of the spin parameter increases, the value of the proportionality function decreases.
\subsection{BTZ BH}\label{BTZ-BH}
\begin{figure}[ht!]
\begin{center}
\includegraphics[width=0.45\textwidth]{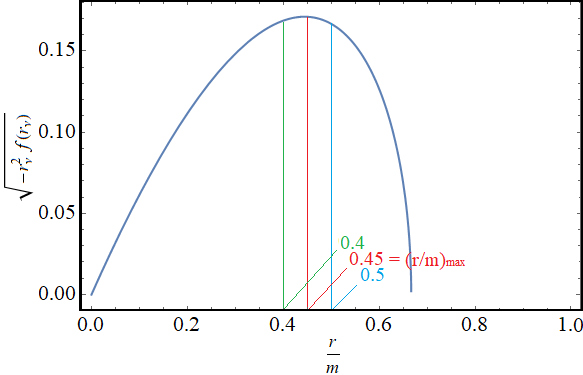}
\caption{The plot of maximization factor $\sqrt{-r_v ^2f(r_v)}$  vs $\frac{r}{m}$ for the rotating BTZ BH shows the maximal hyper-surface located at $0.45m$ with \quad $J=0.5$ \citep{Ali:2020qkb}}.
\label{image-5}
\end{center}
\end{figure}

BTZ BH is a $(2+1)-$dimensional solution of Einstein-Maxwell equations having negative cosmological constant and constant negative curvature \citep{Banados:1992wn, Banados:1992gq, Birmingham:2001dt}. In Ref. \citep{Carlip:1995qv}, it is stated that BTZ BH shares many classical and quantum properties
with the $(3+1)-$ dimensional BH systems. The metric of $(2+1)$-dimensional rotating BTZ BH is defined in Eq. (\ref{metric}), with the lapse function and angular shift are
\begin{equation}
f(r)=-m^2+\frac{r^2}{l^2}+\frac{J^2}{4 r^2},
\end{equation}  
here, we can define $\Lambda=-\frac{1}{l^2}$ as the cosmological constant, $j, m, l$ are the azimuthal angular momentum, AdS mass and AdS radius and $N^\phi (r)$ is the shift function corresponding to angular velocity $\Omega (r)$ and $\phi$ is the period in the range of $0<\phi<2\pi$. For the symmetry the metric of rotating BTZ BHs needs an azimuthal symmetry so that the angular momentum remain conserved under a coordinate transformation. The mass $m$ and Hawking entropy $S_{BH}$ of BTZ rotating BH at the horizon are defined as \citep{Banados:1992wn, Carlip:1995qv}
$$m= \frac{r^2}{l^2}+\frac{J^2}{4 r^2}, \quad S_{BH}=2\pi r_+,$$
At the horizon
\begin{equation}\label{radius}
f(r)=0 \qquad \Rightarrow \qquad r_\pm =\sqrt{\frac{l^2 m}{2}  \left(1\pm X\right)},
\end{equation}
with
$$X=\sqrt{1-\left(\frac{J}{l m}\right)^2}$$
 According to this Eq. (\ref{radius}), the singularity coordinates at  $\frac{J}{l m}=1$. Using the lapse function, one can easily define the surface gravity

$$\kappa=\frac{1}{2}\frac{\partial f(r_+)}{\partial r}=\frac{\sqrt{2}mX}{\sqrt{l^2m(1+X)}} ,$$
and the horizon temperature
\begin{equation}\label{BTZTemp}
T=\frac{mX}{\pi\sqrt{2l^2m(1+X)}},
\end{equation}
The dragging coordinate transformation could help to avoid the dragging effect of the BH so, one should introduce dragging coordinate transformation as

\begin{equation}\label{angularvelocity}
d\phi=-N^{\phi}dt=\frac{J}{2r_+ ^2}dt=\Omega dt,
\end{equation}
This definition of angular velocity is only satisfied in the case of $(2+1)-$dimensional space-time. So, we can investigate the total number of quantum states in the interior of the proposed scalar field. For defining the hyper-surface, the following form of the Eddington Finkelstein coordinates can be used

\begin{equation}\label{Eddmet}
ds^2=\left(-f(r)\dot{v}^2+2 \dot{v} \dot{r}\right) d\lambda^2+r^2\left(N^{\phi}dt+d\phi\right)^2,
\end{equation}
In the case of  axially symmetric BH, there will be negligible effects on the BH horizon, thus the deformative forces will be less effective in the interior of the at $r=r_v$ BH. Summarizing the results, we can write the general form of the BH interior volume as \citep{Ali:2020qkb} 
\begin{equation}
    V_{CR}=\int _0 ^{2\pi} d\phi \int \sqrt{r^2(-f(r)\dot{v}^2+2 \dot{v} \dot{r})}d\lambda=2\pi v \sqrt{-r_v ^2f(r_v)},
\end{equation}
 By the maximization of the factor $\sqrt{-r_v ^2 f(r_v)}$ one can get the maximal hyper-surface
 \begin{equation}
     r_c=\frac{\sqrt{l^2 m \left(\sqrt{3 X^2+1}+2\right)}}{\sqrt{6}},
 \end{equation}
and hence, the interior volume obtained as 
\begin{equation}
    V_{CR} =\frac{v\pi}{3} \sqrt{l^2 m^2 \left(3 X^2+\sqrt{3 X^2+1}+7\right)-9J^2},
\end{equation}
with a numerical position at $r_v=0.45$ as shown in Fig. (\ref{image-5}). 

\section{Entropy Variation and Evaporation}\label{ent-variation}

Using Eq. (\ref{Ent1}), for the rotating BTZ BH the interior entropy of BTZ BH is obtained as
\begin{equation}\label{BTZEnt}
S_{\sum}=\frac{3\zeta (3)}{2\beta^2}\sqrt{-r_v ^2f(r_v)}v,
\end{equation}
From this Eq. (\ref{Ent1}) the entropy of BTZ BH is also proportional to $v$ and this feature may affect the statistical quantities in its interior. To see this, we considered two assumptions to get an evaluation relation between quantum modes entropy and Bekenstein-Hawking entropy. These assumptions are:

\begin{itemize}
\item {BH radiation as black body radiations: So, in $(2+1)$ dimension space-time of rotating BTZ BH, the Boltzmann law is 
\begin{equation}
\frac{dm}{dv}=-\sigma A T^3 \Rightarrow dv=-\frac{\beta^3 \gamma }{ A }dm,
\end{equation}
here $A=\pi  l \sqrt{2 m (X+1)}$ and $\beta$  are the area and the inverse temperature at the event horizon of BTZ BH respectively.}
\item {The BH radiation emission is a quasi-static process (so slow) i.e. $\frac{dm}{dv}<<1 \Rightarrow m<<v$ but Hawking's temperature varies continuously.}
\end{itemize}

So, considering these assumptions with Stefan Boltzmann's law by using the values of $A$ and $\beta$, the differential form of quantum mode entropy is obtained as
\begin{equation}\label{Ent3}
dS_{\sum}=-\frac{3\zeta(3)\gamma \sqrt{-r_v ^2f(r_v)}}{2} \left(\frac{\beta}{ A }\right)dm,
\end{equation}
For a spherically symmetric rotating BH the first law of BH thermodynamics is \citep{Bardeen:1973gs}
\begin{equation}\label{1stlawJ}
{dm}=\frac{d{S}_{BH}}{\beta}+\Omega_{H} dJ,
\end{equation}
here $S_{BH}$ is the Hawking entropy also called horizon entropy. As a conserved quantity, the distortion of angular momentum at the horizon will be small, and at $r=r_v$ it will be negligibly small. So, for onward discussion, the effects of angular momentum will be ignored so, Eq. (\ref{1stlawJ}) becomes
\begin{equation}\label{1^{st}_law}
{dm}=\frac{d{S}_{BH}}{\beta},
\end{equation}

\begin{figure}
\begin{center}
\includegraphics[width=0.45\textwidth]{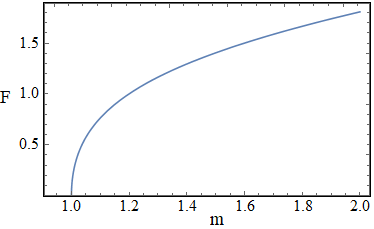}
\caption{A plot of $F(m )$ vs. mass $m$ of rotating BTZ BH \citep{Ali:2020qkb}.}
\label{image-3}
\end{center}
\end{figure}

Using Eq.  (\ref{1^{st}_law}) in Eq.   (\ref{Ent3}), the relation between the interior and exterior entropy of a BH is obtained as
\begin{equation}
dS_{\sum}=-\frac{3\zeta(3)\gamma \sqrt{-r_v ^2f(r_v)}}{2} \left(\frac{dS_{BH}}{ A }\right),
\end{equation}
here, $\zeta$ is the zeta function. This equation is a direct relationship between the two types of entropy. Since the BH's interior entropy is directly related to the interior volume so, by maximizing $\sqrt{-r_v ^2f(r_v)}$, one can maximize the interior entropy. On the other hand at $r=r_v$, BH will have a constant area $A$. Both area and entropy are mass-dependent. This means that this relation between the interior and exterior entropy of a BH is a function of $m$, that is
\begin{equation}\label{proprela}
 dS_{\sum}=-\frac{4 \gamma \zeta (3)}{3 \pi }F(m)d{S}_{\text{BH}},
\end{equation}
here the proportionality function $F(m)$ is 
\begin{equation}\label{Proprel}
F(m)=\frac{\sqrt{m \left(3 X^2+\sqrt{3 X^2+1}-1\right)}}{\sqrt{X+1}},
\end{equation}
The proportional function vs. mass of the BH is plotted in Fig. (\ref{image-3}). This plot represents the power function of some variables between $0$ and $1$. We can see from this curve that the BTZ BH mass $(m)$ grows from $1.0$, and the slope of the curve also increases as the proportionality function increases. At the start point, the BTZ BH mass seems constant with some increase in evolution function. After that, the mass gradually increases, and gaining some mass limits the evolution relation grows with an increase in BH mass uniformly without any deviation, also see Ref. \citep{Ali:2020olc}.

\section{Nature and Probability of Hawking radiation corresponding to Hawking Entropy} \label{natur-radiation}

As in the above discussion, we treated an entangled relation between the interior and exterior entropy for understanding the concept of BH evaporation so, it will be good to understand the nature of radiation emitting from the BH horizon. This relation between BH evaporation and entropy could be the basis for understanding the puzzle of the information loss paradox which is expected to give a detailed theory of quantum gravity \citep{Hawking:2005kf}.

A BH emits Hawking radiation and evaporates but in practice observing them directly from an astrophysical BH is quite difficult. Many ideas are presented to understand the nature and features of these radiations \citep{Wald:1975kc, Hawking:1974rv, Hawking:1975vcx, Brout:1995wp, Michel:2014zsa, Robertson:2012ku, Unruh:1976db, MunozdeNova:2018fxv, Shi:2021nkx} but none of them could explain the proper mechanism for the emission of Hawking radiation.  
According to Hawking, these radiations are one of the quantum features of BH which can be understood by quantum tunneling across the BH horizon and retrieved by many authors e.g., \citep{Wald:1999vt, Vanzo:2011wq, Deng:2016qua}. Considering these tactics for the BH radiation, one can understand the BH interior.  As the entropy of a BH is directly related to its horizon area so, the question can be raised that \textit{"What happens to its entropy after losing energy?".} Once the BH starts evaporating, the Hawking radiation starts escaping from the BH horizon. The quantum mode entropy of this region will no longer be zero due to the entanglement between the interior and exterior quantum modes. This entropy continues to grow as the BH evaporates \citep{Almheiri:2020cfm}. In our discussion, we considered the BH emission rate as a quasi-static process, where the emission process is slow enough and the BH has a variable temperature. As long as the temperature increases, entropy increases, and hence evaporation \footnote{In contrast to this statement, as the BH radiation increases and finally evaporates. It loses all of its information \citep{Hawking:2005kf}. At this stage, it violates the basic Quantum mechanics (QM): that the information of a system must be conserved. In parallel with quantum gravity, one must consider Hawking radiation as the quantum tunneling of particles from the BH horizon \citep{Kraus:1994fh, Parikh:1999mf, Iso:2006ut} i.e., "a virtual particle pair (particle and antiparticle) is created just outside the BH horizon, the antiparticle (negative energy particle) can tunnel through the BH horizon by a process similar to QM tunneling, whereas the particle (positive energy particle) is ejected into spatial infinity. Conversely, for the particle created inside the horizon" \citep{Sakalli:2014sea}. In each case, BH information flows with positive energy particle. }.

One can find the nature of radiations created during the creation-annihilation process of matter by using the the KG equation \citep{Parikh:1999mf, Kraus:1994fh, Iso:2006ut, Sakalli:2014sea}
\begin{equation}
\left(\partial _{\mu }\left(\sqrt{-g}g^{\mu v}\partial _v\right)+m\right)\psi =0,
\end{equation}
where $m$ is the mass of the particle created. In the case of $4-$dimensional space-time coordinates, expanding the above equation gives
\begin{equation}
    \begin{aligned}
        \frac{\partial}{\partial t}\left(\sqrt{-g}g^{\text{tt}}\frac{\partial\psi}{\partial t}\right) +\partial _r\left(\sqrt{-g}g^{\text{rr}}\partial _r\right)+\partial _{\phi }\left(\sqrt{-g}g^{\phi \phi }\partial _{\phi }\right)-\frac{\sqrt{-g} m^2 \psi }{\hbar ^2} \\ =0
    \end{aligned}
\end{equation}

\begin{equation}
\frac{\partial }{\partial t}\left(r g^{\text{tt}} \frac{\partial \psi }{\partial t}\right)+\frac{\partial }{\partial r}\left(\text{rg}^{\text{rr}} \frac{\partial \psi }{\partial r}\right)+\frac{\partial }{\partial \phi }\left(r g^{\phi \phi } \frac{\partial \psi }{\partial \phi }\right)-\frac{m^2 r \psi }{\hbar ^2}=0,
\end{equation}

here $\psi =e^{-i E t}e^{{iI}(r,\phi )}$ so, we get
\begin{equation}
\frac{1}{f(r)}\left(\frac{\partial I}{\partial t}\right)^2-f(r) \left(\frac{\partial I}{\partial r}\right)^2-\frac{1}{r^2}\left(\frac{\partial I}{\partial \phi }\right)^2-m^2=0,
\end{equation}
Applying the method of separating variables with classical action $I(t, r, \phi )=-E t+W(r)+L\phi +c$. So, from the above equation, we get
$$f(r)\left(r \frac{\partial W(r)}{\partial r}\right)^2=\frac{E^2}{f(r)}-\frac{L^2}{r^2}-m^2=0$$
\begin{equation}\label{W}
 \Rightarrow W(r)=\pm \int \frac{\sqrt{E^2-f(r) \left(\left(\frac{L}{r}\right)^2+m^2\right)}}{f(r)} \, dr,
\end{equation}
Here, the $\pm$ signs point to the scalar particles moving away or toward the event horizon (emission/absorption). As $f(r)=0\Rightarrow r=r_h$, one can solve the above equation by using the residue theorem and expanding $f(r_h)$ by the Taylor series, we get 
\begin{equation}
f(r)=f(r)+\left(r_h-r\right) f'(r)+O\left(\left(r_h-r\right)^2\right)\approx \left(r_h-r\right) f'(r),
\end{equation}
So, from above Eq. (\ref{W}), we get the distribution of particles as 
\begin{equation}
W(r)=\pm \frac{\widetilde{E}}{f'(r)}\int \frac{1}{\left(r_h-r\right) } dr,
\end{equation}
here $\widetilde{E}$ is the modified energy. Solving the Residue problem, we get
\begin{equation}
W(r)=\pm \frac{i\pi \widetilde{E}}{f'(r)},
\end{equation}
from this the probabilities of emission/leaving and absorption/entering of particles on  the horizon is 
\begin{equation}\label{absorption}
\Gamma_{+}=e^{-\frac{2}{\hbar}ImI}=e^{-\frac{2}{\hbar}(ImW_+ +Imc)},
\end{equation}
\begin{equation}\label{emission}
\Gamma_{-}=e^{-\frac{2}{\hbar}ImI}=e^{-\frac{2}{\hbar}(ImW_- +Imc)},
\end{equation}
An object in the vicinity of the BH horizon will be swallowed by it. So, the absorption probability in Eq. (\ref{absorption}) could be normalized to unity. We can do this by considering $Imc=-ImW_-$. We also know that $ImW_+=-ImW_-$. So, from the above two Eqs. (\ref{absorption}) and (\ref{emission}), we get
\begin{equation}
\Gamma_{+}=e^{-\frac{4}{\hbar}ImW_+}=e^{-\frac{4\pi}{f'(r)} \frac{\widetilde{E}}{\hbar}},
\end{equation}
It means that during the tunneling process of particles from the BH horizon, one can't distinguish the particles. Using the value of $f'(r)$, we have
\begin{equation}\label{bolt}
\Gamma_+=e^{-\beta \frac{\widetilde{E}}{\hbar}}=e^{-\beta \omega},
\end{equation}
Here, we used $\widetilde{E}=\hbar \omega$ and $\beta$ is the inverse Hawking temperature of BH. The above Eq. (\ref{bolt}) is the Boltzmann distribution formula. Here the leading order term $e^{-\beta \omega}$ is the Boltzmann factor for emitted radiation. From this Eq. (\ref{bolt}), we can easily obtain the horizon temperature and entropy \citep{Li:2020qqa, Sakalli:2015raa}.
\section{Summary}
The study of the BH interior could reveal many facts that need to be understood about the BH. After the CR work \citep{Christodoulou:2014yia} in 2015, the notion of BH interior volume got a great attraction from many authors and great work has been done on this topic. Starting from BH's interior volume, BH's entropy, evaporation, and the information paradox issue are discussed with groundbreaking results. Many valuable results have been found so far.  This paper reviews all related works on BH's interior volume, entropy, evaporation, and a possible solution to the information paradox by using BHs in different spacetime dimensions.

From Parikh's work (the BH volume remains constant over time) to the perspective of Christodoulou and Rovelli (the volume of a BH varies with time), we employed a unique method to determine the volume of the BH interior. The concept of BH's interior volume seems different from that bound by a sphere in flat $3d$ spacetime. The interior volume bound by a $2-$sphere immersed in curved space-time is the volume of the largest space-like spherically symmetric hyper-surface bounded by two-sphere $S$. This means that the BH bounds a maximum volume equal to the volume bounded by the largest $3d$ hyper-surface and is found proportional to $v$. Following their investigations, the quantum mode entropy in the scalar field is investigated directly proportional to $v$ by Baocheng Zhang \citep{Zhang:2015gda}. From these time-dependent relations of BH interior volume and entropy with $v$, it is greatly possible to affect the statistical quantities with a small change in interior volume or entropy in the interior of BH. It is a great step to follow up for probing the idea of information paradox. After investigating the interior volume and entropy, we determined an entanglement relation between the interior and exterior entropy to understand the change in BH thermodynamical quantities and the evaporation status with these changes. For this purpose, two assumptions of BH radiation as black body radiation and the emission process of radiation as a quasi-static were introduced. The $1^{st}$ assumption led us to use the Boltzmann law and the $2^{nd}$ assumption guaranteed us to use the differential form investigation up to the quantum level for an infinitely small interval of time. Using these assumptions, we got the differential form of the interior and exterior entropy. By comparing the two types of entropy, we obtained proportional relationships. It shows some important features consistent with Hawking's investigation of BH evaporation. By applying this technique in different BH space-times, we get an extended confirmation of our results as given in the tables (\ref{tab:1}) and (\ref{tab:2}). An exemplary review is also made for Schawarzschild, Kerr, and BTZ-type BH space-times. 

Using the assumption of quasi-static emission, we also investigated the emission probability and nature of BH radiation that satisfies the Boltzmann distribution. The largest number of quantum states $\left(N_{BH}\propto \frac{1}{\Gamma_+}\right)$, that a BH could reside is related to $S_{BH}$ as 
 \begin{equation}
     N_{BH}=e^{S_{BH}}
 \end{equation}
But during the evaporation process, the Bekenstein and Hawking entropy gradually reduces as also clear from Eq. (\ref{Qmode}), and the number of quantum states grows with the expansion of BH's volume, thus the interior entropy of the scalar field also increases. due to this reason, at the final stage of evaporation (stop point) the number of quantum states in the BH's interior is much more than the exterior quantum modes of BH. Hence this investigation confirms the result of \citep{Christodoulou:2014yia} and one can claim that the large interior volume will have more space to store information even if the horizon shrinks to a small size (Remnant BH) \citep{Chen:2014jwq}. This is the main point of our investigation to claim a large interior of the BH to store information.

Finally, we can suggest that work is needed on these topics to more deeply understand BH physics and solve the issue of the information loss paradox. In this regard, we could consider the modified gravity theories in our future work. 

\section{Acknowledgments}

I sincerely thank and appreciate the editor and anonymous referee's effort for careful reading and helpful suggestions on this manuscript. Their suggestions significantly improved the work done in this paper. The main credit for this work also goes to my kind supervisor Prof. Liu Wen-Biao and my lovely friends X. Y. Wang and Ming Zhang who helped me throughout my PHD duration. This work was supported by the National Key R\&D Program of China (Grant No. 2023YFA1607902), the National Natural Science Foundation of China under grants 12173031 and 12221003, and the science research grants from the China Manned Space Project.

\section{Appendices}
\subsection{Appendix A.}\label{A}
Consider a hyper-surface at a late advanced time $v$. Let us extend it from the event horizon with a constant radius $r$ in the interior of a BH. When it reaches the point where $v$ deviates give the maximal advance time and the hyper-surface at this point will bind the largest interior volume. The decomposed form of an arbitrary vector lying on a hyper-surface can be written
\begin{equation}
k=Zn^a+Z^a \qquad \Rightarrow \qquad n_a=-Z\nabla_a{r}
\end{equation}
where $Z$ and $Z^a$ are the Lapse and shift functions respectively, and $n^a$ is the co-vector with $\nabla_a{r}$ being the normal co-vector. For a space-like hyper-surface, the normal vector can  be written as 
\begin{equation}
n_a n^a=-1 \qquad \Rightarrow \qquad Z^2g^{ab}dr_a dr_b=-1,
 \end{equation}
Using the above equation, we can write as $Z^2=-g^{-rr}$. Now as the determinant of an induced metric on hyper-surface at constant radius $r$ is 
\begin{equation}
|h|=Z^{-2}(g)=g^{rr}g \qquad\Rightarrow \qquad |h|^2=-\Delta\rho^2 sin^2\theta,
\end{equation}
Now the volume for a rotating BH, the interior volume can be defined as 
\begin{equation}
V_{\sum}=\int^v\sqrt{|h|}dv d\theta d\phi=\int^v \sqrt{-\Delta}\rho sin\theta dv d\theta d\phi,
\end{equation}
This directly confirms the BH interior volume as in Ref. \citep{Bengtsson:2015zda}.

\subsection{Appendix B.}\label{B}
 The Klein-Gordon equation in curved space-time is,
\begin{equation}
    \frac{1}{\sqrt{-g}}{{\partial_ \mu }\left(\sqrt{-g} g^{\mu v } \partial_v \Phi \right)}=0,
\end{equation}
Expanding this equation in $(T,\lambda, \theta, \phi)$ and using
$$\partial\Phi_v=\partial_ T  \Phi+\partial_{\lambda} \Phi+\partial_ {\theta} \Phi+\partial_ \phi \Phi,$$
First solving $(\partial_ T  \Phi, \partial_{\lambda} \Phi, \partial_ {\theta} \Phi, \partial_ \phi \Phi)$, we get
\begin{equation}
    \begin{aligned}
        \partial_ T  \Phi=\partial_ T (e^{\text{-iET}}e^{\text{iI}(\lambda ,\theta ,\phi )})=-i E\Phi,\\ \partial_{\lambda} \Phi=\partial_{\lambda} (e^{\text{iET}}e^{\text{iI}(\lambda ,\theta ,\phi )})=i I\partial_{\lambda} \Phi,\\ \partial_ {\theta} \Phi=\partial_ {\theta} (e^{\text{iET}}e^{\text{iI}(\lambda ,\theta ,\phi )})=i I\partial_ {\theta} \Phi, \\ \partial_ \phi \Phi)=\partial_ \phi (e^{\text{iET}}e^{\text{iI}(\lambda ,\theta ,\phi )})=i I \partial_ \phi \Phi,
    \end{aligned}
\end{equation}

So, from above we can write the scalar field as 
$$\partial_v \Phi=i (E^2+\partial_{\lambda} I+\partial_ {\theta} I+\partial_ \phi I)\Phi,$$
The first part of the above general equation can be written as 
\begin{equation*}
    \begin{aligned}
        \frac{1}{\sqrt{-g}}{{\partial_ T }\left(\sqrt{-g} g^{T v } \partial_v \Phi \right)} \\
=\frac{1}{\sqrt{-g}}{{\partial_ T }\left ( \sqrt{-g}\left ( g^{TT} +g^{T\lambda}+g^{T\theta}+g^{T\phi}\right )\partial_v \Phi  \right )},
    \end{aligned}
\end{equation*}
\begin{small}
    \begin{equation*}
    \begin{aligned}
        \frac{1}{\sqrt{-g}}{{\partial_ T }\left(\sqrt{-g} g^{T v } \partial_v \Phi \right)}\\ =\frac{1}{\sqrt{-g}}{{\partial_ T }\left ( \sqrt{-g}\left ( g^{TT} +g^{T\lambda}+g^{T\theta}+g^{T\phi}\right )\times i (E^2+\partial_{\lambda} I+\partial_ {\theta} I+\partial_ \phi I)\Phi \right)},
    \end{aligned}
\end{equation*}
\begin{equation*}
\begin{aligned}
    \frac{1}{\sqrt{-g}}{{\partial_ T }\left(\sqrt{-g} g^{T v } \partial_v \Phi \right)} \\ =\frac{1}{\sqrt{-g}}{{\partial_ T }\left ( \sqrt{-g}\left ( g^{TT} +g^{T\lambda}+g^{T\theta}+g^{T\phi}\right ) \times i (E^2+\partial_{\lambda} I+\partial_ {\theta} I+\partial_ \phi I)\Phi \right)},
\end{aligned}
\end{equation*}
\end{small}
where for the metric Eq. (\ref{EDmet+T}), the coordinates are $g^{TT}=-1, g^{T\lambda}=g^{T\theta}=g^{T\phi}=0$

\begin{equation}
    \begin{aligned}
        \frac{1}{\sqrt{-g}}{{\partial_ T }\left(\sqrt{-g} g^{T v } \partial_v \Phi \right)} \\ =\frac{1}{\sqrt{-g}}{{\partial_ T }\left ( \sqrt{-g}\left ( -1 +0+0+0\right )\times i (E^2+\partial_{\lambda} I+\partial_ {\theta} I+\partial_ \phi I)\Phi \right)},
    \end{aligned}
\end{equation}
or we can write as 
$$\frac{1}{\sqrt{-g}}{{\partial_ T }\left(\sqrt{-g} g^{T v } \partial_v \Phi \right)}=- E^2\Phi,$$
Similarly, the other parts of the general equation are

\begin{small}
    \begin{equation}
    \begin{aligned}
        \frac{1}{\sqrt{-g}}{{\partial_ \lambda }\left(\sqrt{-g} g^{\lambda  v } \partial_v \Phi \right)} \\ =\frac{1}{\sqrt{-g}}{{\partial_ \lambda }\left ( \sqrt{-g}\left ( g^{\lambda T} +g^{\lambda\lambda}+g^{\lambda\theta}+g^{\lambda\phi}\right )\times i (E^2+\partial_{\lambda} I+\partial_ {\theta} I+\partial_ \phi I)\Phi \right)},
    \end{aligned}
\end{equation}
\end{small}
Here $g^{\lambda T}=0, g^{\lambda\lambda}=\frac{1}{\left(-f(r) \dot{v }^2+2 \dot{v } \dot{r}\right)} ,g^{\lambda\theta}=0, g^{\lambda\phi}=0$ so,
$$\frac{1}{\sqrt{-g}}{{\partial_ \lambda }\left(\sqrt{-g} g^{\lambda  v } \partial_v \Phi \right)}=\frac{1}{\left(-f(r) \dot{v }^2+2 \dot{v } \dot{r}\right)} \partial_ \lambda ^2 I \Phi,$$

\begin{small}
\begin{equation}
    \begin{aligned}
        \frac{1}{\sqrt{-g}}{{\partial_ \theta }\left(\sqrt{-g} g^{\theta  v } \partial_v \Phi \right)} \\ =\frac{1}{\sqrt{-g}}{{\partial_ \theta }\left ( \sqrt{-g}\left ( g^{\theta T} +g^{\theta\lambda}+g^{\theta\theta}+g^{\theta\phi}\right )\times i (E^2+\partial_{\lambda} I+\partial_ {\theta} I+\partial_ \phi I)\Phi \right)}
    \end{aligned}
\end{equation}
\end{small}

Here $$g^{\theta T}=0, g^{\theta\lambda}=0 ,g^{\theta\theta}=\frac{1}{r^2}, g^{\theta\phi}=0$$ so,
$$=\frac{1}{\sqrt{-g}}{{\partial_ \theta }\left ( \sqrt{-g}\left ( 0 +0+\frac{1}{r^2}+0\right )\times i (E^2+\partial_{\lambda} I+\partial_ {\theta} I+\partial_ \phi I)\Phi \right)}$$ 
$$\frac{1}{\sqrt{-g}}{{\partial_ \theta }\left(\sqrt{-g} g^{\theta  v } \partial_v \Phi \right)}=\frac{1}{r^2}\partial^2 _ {\theta}I\Phi$$

and 

\begin{small}
\begin{equation}
    \begin{aligned}
       \frac{1}{\sqrt{-g}}{{\partial_ \phi }\left(\sqrt{-g} g^{\phi  v } \partial_v \Phi \right)} \\ =\frac{1}{\sqrt{-g}}{{\partial_ \phi }\left ( \sqrt{-g}\left ( g^{\phi T} +g^{\phi\lambda}+g^{\phi\theta}+g^{\phi\phi}\right )\times i (E^2+\partial_{\lambda} I+\partial_ {\theta} I+\partial_ \phi I)\Phi \right)},
    \end{aligned}
\end{equation}
\end{small}

here $g^{\phi T}=0, g^{\phi\lambda}=0 ,g^{\phi\theta}=0, g^{\phi\phi}=\frac{1}{r^2sin^2 \theta}$ so,

\begin{small}
    \begin{equation}
    \begin{aligned}
        =\frac{1}{\sqrt{-g}}{{\partial_ \phi }\left ( \sqrt{-g}\left ( 0 +0+0+\frac{1}{r^2 sin^2 \theta}\right )\times i (E^2+\partial_{\lambda} I+\partial_ {\theta} I+\partial_ \phi I)\Phi \right)},
    \end{aligned}
\end{equation}

\end{small}

$$\Rightarrow \frac{1}{\sqrt{-g}}{{\partial_ \phi }\left(\sqrt{-g} g^{\phi v } \partial_v \Phi \right)}=\frac{1}{r^2 sin^2 \theta}\partial^2 _ {\phi}I\Phi,$$
using the above results, the general solution of the Klein-Gordon equation in curved space-time for $(T,\lambda, \theta, \phi)$ can be written as 
$$- E^2\Phi+\frac{1}{\left(-f(r) \dot{v }^2+2 \dot{v } \dot{r}\right)} \partial_ \lambda ^2 I \Phi+\frac{1}{r^2}\partial^2 _ {\theta}I\Phi+\frac{1}{r^2 sin^2 \theta}\partial^2 _ {\phi}I\Phi=0,$$
using $\partial_ \lambda ^2 I=p_\lambda ^2, \partial^2 _ {\theta}I=p_\theta ^2, \partial^2 _ {\phi}I=p_\phi ^2$, we gets 

\begin{equation}
    -E^2+\frac{1}{\left(-f(r) \dot{v }^2+2 \dot{v } \dot{r}\right)} p_\lambda ^2 +\frac{1}{r^2}p_\theta ^2+\frac{1}{r^2 sin^2 \theta}p_\phi ^2=0,
\end{equation}

Which is the Eq. (\ref{eqofenergy}). One can easily deduce these investigations for the lower dimensional BH and obtain the EOM.

\bibliographystyle{elsarticle-harv} 
\bibliography{bibliography}
\end{document}